\documentclass[11pt]{article}

\usepackage{amsthm,amsmath}
\RequirePackage[numbers]{natbib}
\RequirePackage[colorlinks,citecolor=blue,urlcolor=blue]{hyperref}

\usepackage{algorithm}
\usepackage{enumerate}
\usepackage{ifthen}
\usepackage[utf8]{inputenc}
\usepackage{framed}

\usepackage[utf8]{inputenc}
\usepackage{makeidx}
\usepackage{amsfonts}
\usepackage{amssymb}
\usepackage{booktabs}

\usepackage{gensymb}
\usepackage{tabularx}
\usepackage{float}
\usepackage{caption}
\usepackage[labelformat=simple]{subcaption}
\usepackage{bbm}
\usepackage{lipsum}
\usepackage{enumitem}
\usepackage{hyperref}
\usepackage[english]{babel} 
\usepackage{bm} 
\usepackage[T1]{fontenc}	 
\usepackage{lmodern}
\usepackage{amsthm} 
\usepackage{graphicx} 
\usepackage{booktabs} 
\usepackage{multicol} 
\usepackage[toc,page]{appendix}

\newcommand{\footremember}[2]{%
	\footnote{#2}
	\newcounter{#1}
	\setcounter{#1}{\value{footnote}}%
}

\DeclareMathOperator*{\argmax}{arg\,max}

\def\C {\,|\:}

\allowdisplaybreaks
\numberwithin{equation}{section}

\theoremstyle{plain}

\allowdisplaybreaks

\begin{document}

\hypersetup{linkcolor=blue}

\date{\today}

\author{Ray Bai\footremember{UofSC}{Department of Statistics,  University of South Carolina, Columbia, SC 29208. Email: \href{mailto:RBAI@mailbox.sc.edu}{\tt RBAI@mailbox.sc.edu}}, 
	 Veronika Ro\v{c}kov\'{a}\footremember{Booth}{Booth School of Business, University of Chicago, Chicago, IL, 60637. E-mail: \href{mailto:Veronika.Rockova@chicagobooth.edu}{\tt Veronika.Rockova@chicagobooth.edu}}, Edward I. George\footremember{Wharton}{Department of Statistics, The Wharton School, University of Pennsylvania, Philadelphia, PA, 19104.  E-mail: \href{mailto:edgeorge@wharton.upenn.edu}{\tt edgeorge@wharton.upenn.edu} } }

\title{Spike-and-Slab Meets LASSO: A Review of the Spike-and-Slab LASSO  
}

\maketitle

\begin{abstract}
\noindent High-dimensional data sets have become ubiquitous in the past few decades, often with many more covariates than observations. In the frequentist setting, penalized likelihood methods are the most popular approach for variable selection and estimation in high-dimensional data. In the Bayesian framework, spike-and-slab methods are commonly used as probabilistic constructs for high-dimensional modeling. Within the context of linear regression, \cite{RockovaGeorge2018} introduced the spike-and-slab LASSO (SSL), {  an approach based on a prior which provides a continuum between the penalized likelihood LASSO and the Bayesian point-mass spike-and-slab formulations}. Since its inception, the spike-and-slab LASSO has been extended to a variety of contexts, including generalized linear models, factor analysis, graphical models, and nonparametric regression. The goal of this paper is to survey the landscape surrounding spike-and-slab LASSO methodology. First we elucidate the attractive properties and the computational tractability of SSL priors in high dimensions. We then review methodological developments of the SSL and outline several theoretical developments. We illustrate the methodology on both simulated and real datasets.  

\end{abstract}

\section{Introduction} \label{Intro}

High-dimensional data are now routinely analyzed. In these settings, one often wants to impose a low-dimensional structure such as sparsity. For example, in astronomy and other image processing contexts, there may be thousands of noisy observations of image pixels, but only a small number of these pixels are typically needed to recover the objects of interest \cite{JohnstoneSilverman2004,JohnstoneSilverman2005}. In genetic studies, scientists routinely observe tens of thousands of gene expression data points, but only a few genes may be significantly associated with a phenotype of interest. For example, \cite{WellcomeTrust2007} has confirmed that only seven genes have a non-negligible association with Type I diabetes. Among practitioners, the main objectives in these scenarios are typically: a) identification (or \textit{variable selection}) of the non-negligible variables, and b) \textit{estimation} of their effects.

A well-studied model for sparse recovery in the high-dimensional statistics literature is the normal linear regression model,
\begin{equation} \label{linearregression}
	\boldsymbol{y} = \boldsymbol{X} \boldsymbol{\beta} + \boldsymbol{\varepsilon}, \hspace{.5cm} \boldsymbol{\varepsilon} \sim \mathcal{N}_n (\boldsymbol{0}, \sigma^2 \boldsymbol{I}_n ),
\end{equation}
where $\boldsymbol{y} \in \mathbb{R}^{n}$ is a vector of $n$ responses, $\boldsymbol{X} = [ \boldsymbol{x}_1, \ldots, \boldsymbol{x}_p] \in \mathbb{R}^{n \times p}$ is a design matrix of $p$ potential covariates, $\boldsymbol{\beta} = (\beta_1, \ldots, \beta_p)^T$ is a $p$-dimensional vector of unknown regression coefficients, and $\boldsymbol{\varepsilon}$ is the noise vector. When $p > n$, we often assume that most of the elements in $\boldsymbol{\beta}$ are zero or negligible. Under this setup, there have been a large number of methods proposed for selecting and estimating the active coefficients in $\boldsymbol{\beta}$. In the frequentist framework, penalized likelihood approaches such as the least absolute shrinkage and selection operator (LASSO) \cite{Tibshirani1996} are typically used to achieve sparse recovery for $\boldsymbol{\beta}$. In the Bayesian framework, spike-and-slab priors are a popular approach for sparse modeling of $\boldsymbol{\beta}$. 

The spike-and-slab LASSO (SSL), introduced by \cite{RockovaGeorge2018}, forms a continuum between these penalized likelihood and spike-and-slab constructs. The spike-and-slab LASSO methodology has experienced rapid development in recent years, and its scope now extends well beyond the normal linear regression model \eqref{linearregression}. The purpose of this paper is to offer a timely review of the SSL and its many variants. We first provide a basic review of frequentist and Bayesian approaches to high-dimensional variable selection and estimation under the model \eqref{linearregression}. We then review the spike-and-slab LASSO and provide an overview of its attractive properties and techniques to implement it within the context of normal linear regression. Next, we review the methodological developments of the SSL and some of its theoretical developments. 

\section{Variable selection in high dimensions: Frequentist and Bayesian strategies} \label{FrequentistVsBayes}

We first review the frequentist penalized regression and the Bayesian spike-and-slab frameworks before showing how the spike-and-slab LASSO bridges the gap between them.

\subsection{Penalized likelihood approaches} \label{PenalizedLikelihood}

In the frequentist high-dimensional literature, there have been a variety of penalized likelihood approaches proposed to estimate $\boldsymbol{\beta}$ in \eqref{linearregression}. A variant of the penalized likelihood approach estimates $\boldsymbol{\beta}$ with 
\begin{equation} \label{penalizedlikelihood}
	\widehat{\boldsymbol{\beta}} = \argmax_{\boldsymbol{\beta} \in \mathbb{R}^{p}} - \frac{1}{2} \lVert \boldsymbol{y} - \boldsymbol{X} \boldsymbol{\beta} \rVert_2^2 + \textrm{pen}_{\lambda} (\boldsymbol{\beta}),
\end{equation}
where $\textrm{pen}_{\lambda} (\boldsymbol{\beta})$ is a penalty function indexed by penalty parameter $\lambda$. Most of the literature has focused on penalty functions which are separable, i.e. $\textrm{pen}_{\lambda} (\boldsymbol{\beta}) = \sum_{j=1}^{p} \rho_{\lambda} (\beta_j)$. In particular, the popular least absolute shrinkage and selection operator (LASSO) penalty of \cite{Tibshirani1996} uses the function $\rho_{\lambda} (\beta_j) = - \lambda \lvert \beta_j \rvert$. Besides the LASSO and its many variants \cite{BelloniChernozhukovWang2011,SunZhang2012,ZhangZhang2014,Zou2006,ZouHastie2005}, other popular choices for $\rho_{\lambda} (\cdot)$ include non-concave penalty functions, such as the smoothly clipped absolute deviation (SCAD) penalty \cite{FanLi2001} and the minimax concave penalty (MCP) \cite{Zhang2010}. All of the aforementioned penalties threshold some coefficients to zero, thus enabling them to perform variable selection and estimation simultaneously. In addition, SCAD and MCP also mitigate the well-known estimation bias of the LASSO.

Any penalized likelihood estimator \eqref{penalizedlikelihood} also has a Bayesian interpretation in that it can be seen as a posterior mode under an independent product prior $p(\boldsymbol{\beta} \C \lambda) = \prod_{j=1}^{p} p(\beta_j \C \lambda)$, where $\textrm{pen}_{\lambda} (\boldsymbol{\beta}) = \log p (\boldsymbol{\beta} \C \lambda) = \sum_{j=1}^{p} \log p (\beta_j \C \lambda)$. In particular, the solution to the LASSO is equivalent to the posterior mode under a product of Laplace densities indexed by hyperparameter, $\lambda$:
\begin{align} \label{bayesianLASSO}
	p ( \boldsymbol{\beta} \C \lambda) = \prod_{j=1}^{p} \frac{\lambda}{2} e^{- \lambda \lvert \beta_j \rvert }.
\end{align}
This prior, known as the Bayesian LASSO, was first introduced by \cite{ParkCasella2008}. In \cite{ParkCasella2008}, both fully Bayes and empirical Bayes procedures were developed to tune the hyperparameter $\lambda$ in \eqref{bayesianLASSO}. The fully Bayes approach of placing a prior on $\lambda$, in particular, renders the Bayesian LASSO penalty \textit{non}-separable. Thus, the fully Bayesian LASSO has the added advantage of being able to share information across different coordinates. Despite this benefit, \cite{RockovaGeorge2016Abel} showed that the fully Bayesian LASSO cannot simultaneously adapt to sparsity \textit{and} avoid the estimation bias issue of the original LASSO. In addition, \cite{GhoshTangGhoshChakrabarti2016} showed that the univariate Bayesian LASSO often undershrinks negligible coefficients, while overshrinking large coefficients. Finally,  \cite{CastilloSchmidtHieberVanDerVaart2015} also proved that the posterior under the Bayesian LASSO contracts at a suboptimal rate. In Sections \ref{SpikeAndSlabLASSO}-\ref{Illustration}, we will illustrate how the \textit{spike-and-slab LASSO} mitigates these issues.

In addition to the spike-and-slab LASSO, other alternative priors have also been proposed to overcome the limitations of the Bayesian LASSO \eqref{bayesianLASSO}. These priors, known as global-local shrinkage (GL) priors, place greater mass around zero and have heavier tails than the Bayesian LASSO. Thus, GL priors shrink small coefficients more aggressively towards zero, while their heavy tails prevent overshrinkage of large coefficients. Some examples include the normal-gamma prior \cite{GriffinBrown2010}, the horseshoe prior \cite{CarvalhoPolsonScott2010}, the generalized double Pareto prior \cite{ArmaganDunsonLee2013}, the Dirichlet-Laplace prior \cite{BhattacharyaPatiPillaiDunson2015}, and the normal-beta prime prior \cite{BaiGhosh2021}. We refer the reader to \cite{BhadraDattaPolsonWillard2019} for a detailed review of GL priors.

\subsection{Spike-and-slab priors} \label{sspriors}
In the Bayesian framework, variable selection under the linear model \eqref{linearregression} arises directly from probabilistic considerations and has frequently been carried out through placing spike-and-slab priors on the coefficients of interest. The spike-and-slab prior was first introduced by \cite{MitchellBeauchamp1988} and typically has the following form,
\begin{equation} \label{pointmassspikeandslab}
	\begin{array}{rl}
		p (\boldsymbol{\beta} \C \boldsymbol{\gamma}, \sigma^2 ) = & \displaystyle \prod_{j=1}^{p} \left[ (1-\gamma_j) \delta_0 (\beta_j) + \gamma_j p (\beta_j \C \sigma^2 ) \right], \\
		p (\boldsymbol{\gamma} \C \theta) = & \displaystyle \prod_{j=1}^{p} \theta^{\gamma_j} (1-\theta)^{1-\gamma_j}, \hspace{.5cm} \theta \sim p(\theta), \\
		\sigma^2 \sim & p(\sigma^2),
	\end{array}
\end{equation}
where $\delta_0$ is a point mass at zero used to model the negligible entries (the ``spike''), $p( \beta_j \C \sigma^2)$ is a diffuse and/or heavy-tailed density (rescaled by the variance $\sigma^2$) to model the non-negligible entries (the ``slab''), $\boldsymbol{\gamma}$ is a binary vector that indexes the $2^p$ possible models, and $\theta \in (0,1)$ is a mixing proportion. The error variance $\sigma^2$ is typically endowed with a conjugate inverse gamma prior or an improper Jeffreys prior, $p (\sigma^2) \propto \sigma^{-2}$. With a well-chosen prior on $\theta$, this prior \eqref{pointmassspikeandslab} also automatically favors parsimonious models in high dimensions, thus avoiding the curse of dimensionality. 

The point-mass spike-and-slab prior \eqref{pointmassspikeandslab} is often considered ``theoretically ideal,'' or a ``gold standard'' for sparse Bayesian problems \cite{CarvalhoPolsonScott2009,PolsonSun2019,Rockova2018}. In high dimensions, however, exploring the full posterior over the entire model space using point-mass spike-and-slab priors \eqref{pointmassspikeandslab} can be computationally prohibitive, in large part because of the combinatorial complexity of updating the discrete indicators $\boldsymbol{\gamma}$. There has been some work to mitigate this issue by using either shotgun stochastic search (SSS) \cite{BottoloRichardson2010,HansDobraWest2007} or variational inference (VI) \cite{RaySzabo2020} to quickly identify regions of high posterior probability.

As an alternative to the point-mass spike-and-slab prior, fully continuous spike-and-slab models have been developed. In these continuous variants, the point-mass $\delta_0$ in \eqref{pointmassspikeandslab} is replaced by a continuous density that is heavily concentrated about zero. The first such continuous relaxation was made by \cite{GeorgeMcCulloch1993}, who used a normal density with very small variance for the spike and a normal density with very large variance for the slab. Specifically, the prior for $\boldsymbol{\beta}$ in \cite{GeorgeMcCulloch1993} is
\begin{equation} \label{spikeandslabnormals}
	p(\boldsymbol{\beta} \C \boldsymbol{\gamma}, \sigma^2) = \prod_{j=1}^{p} \left[ (1-\gamma_j) \mathcal{N}(0, \sigma^2 \tau_0^2) + \gamma_j \mathcal{N}(0, \sigma^2 \tau_1^2) \right],
\end{equation} 
where $0 < \tau_0^2 \ll \tau_1^2$. \cite{GeorgeMcCulloch1993} developed a stochastic search variable selection (SSVS) procedure based on posterior sampling with Markov chain Monte Carlo (MCMC) and thresholding the posterior inclusion probabilities, $\Pr (\gamma_j = 1 \C \boldsymbol{y}), j=1, \ldots, p$. In practice, the ``median thresholding'' rule \cite{BarbieriBerger2004}, i.e. $\Pr (\gamma_j = 1 \C \boldsymbol{y}) > 0.5, j=1, \ldots, p$, is often used to perform variable selection.  \cite{IshwaranRao2005, NarisettyHe2014} further extended the model \eqref{spikeandslabnormals} by rescaling the variances $\tau_0^2$ and $\tau_1^2$ with sample size $n$ in order to better control the amount of shrinkage for each individual coefficient. 

To further reduce the computational intensiveness of SSVS, a deterministic optimization procedure called EM variable selection (EMVS) was developed by \cite{RockovaGeorge2014}. The EMVS procedure employs \eqref{spikeandslabnormals} as the prior for $\boldsymbol{\beta}$ and uses an EM algorithm to target the posterior mode for $(\boldsymbol{\beta}, \theta, \sigma )$. (\cite{RockovaGeorge2014} also consider continuous spike-and-slab models where the slab, $\mathcal{N}(0, \sigma^2 \tau_1^2)$, is replaced with a polynomial-tailed density, such as a Student's $t$ or a Cauchy distribution, to prevent overshrinkage of the non-negligible entries in $\boldsymbol{\beta}$).   Compared to SSS and SSVS, EMVS has been shown to more rapidly and reliably identify those sets of higher probability submodels which may be of most interest \cite{RockovaGeorge2014}. Recently, \cite{KimGao2019} proposed a general algorithmic framework for Bayesian variable selection with graph-structured sparsity which subsumes the EMVS algorithm as a special case. Like  SSVS, these algorithms also require thresholding the posterior inclusion probabilities to perform variable selection. Letting $(\boldsymbol{\widehat{\beta}}, \widehat{\theta}, \widehat{\sigma})$ denote the posterior mode for $(\boldsymbol{\beta}, \theta, \sigma)$, \cite{RockovaGeorge2014} recommend using median thresholding, $\Pr(\gamma_j = 1 \C \boldsymbol{\widehat{\beta}}, \widehat{\theta}, \widehat{\sigma}) > 0.5$, for selection.

\section{The spike-and-slab LASSO} \label{SpikeAndSlabLASSO}

Having reviewed the penalized likelihood and spike-and-slab paradigms for sparse modeling in the normal linear regression model \eqref{linearregression}, we are now in a position to review the spike-and-slab LASSO (SSL) of \cite{RockovaGeorge2018}. The SSL forms a bridge between these two parallel developments, thereby combining the strengths of both approaches into a single procedure. Throughout this section and Section \ref{Computing}, we assume that $\boldsymbol{y}$ has been centered at zero to avoid the need for an intercept and that the design matrix $\boldsymbol{X}$ has been centered and standardized so that $\lVert \boldsymbol{x}_j \rVert_2^2 = n$ for all $1 \leq j \leq p$.

\subsection{Prior specification}

\begin{figure}[t!]
	\centering
	\includegraphics[width=0.7\textwidth]{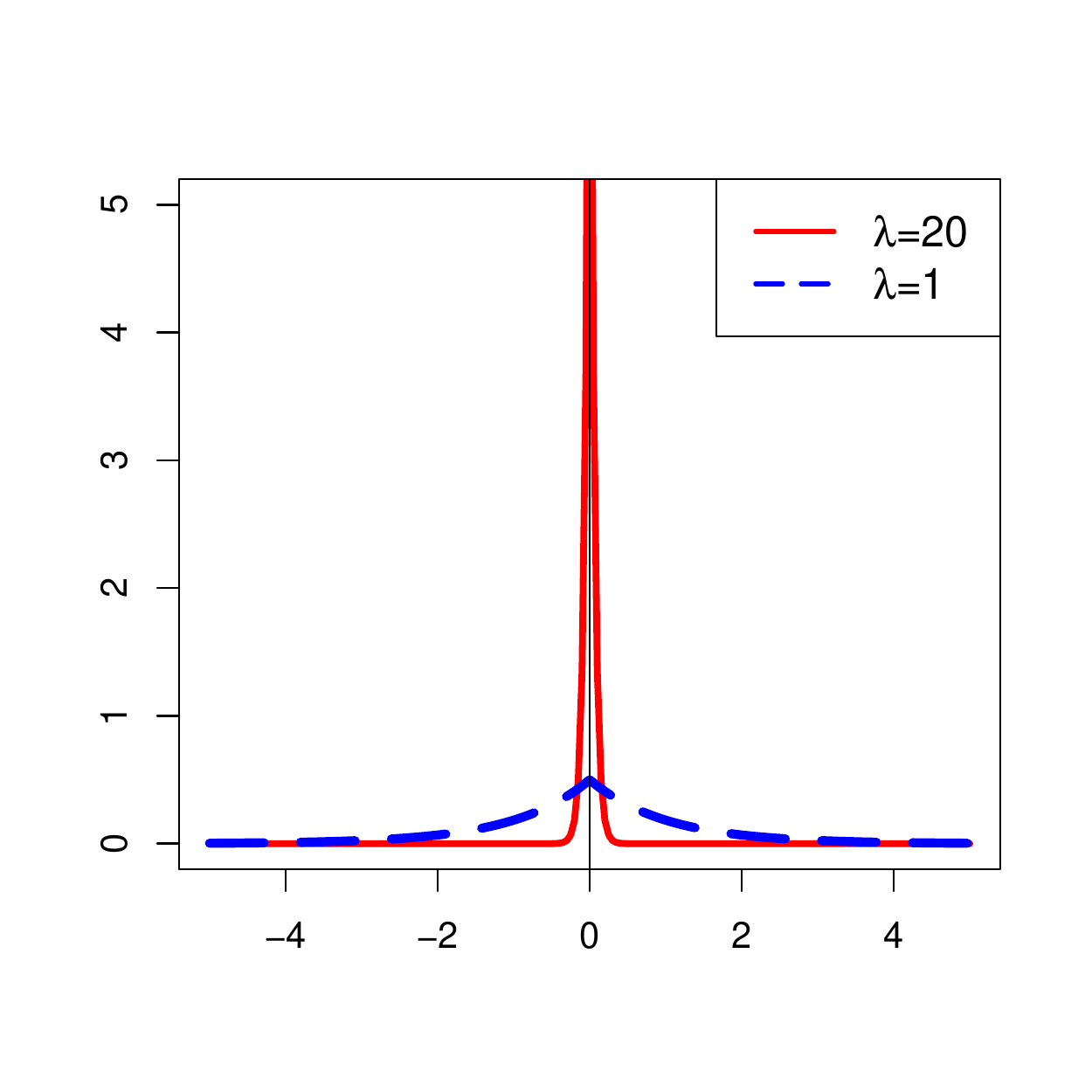} 
	\caption{Plot of the central region for the Laplace density with two different choices of scale parameter. }
	\label{fig:laplace}
\end{figure}

The spike-and-slab LASSO prior is specified as
\begin{equation} \label{SSLprior}
	\begin{array}{rl}
		p(\boldsymbol{\beta}\C\boldsymbol{\gamma} ) = & \displaystyle \prod_{j=1}^{p} \left[ (1-\gamma_j) \psi(\beta_j \C \lambda_0) + \gamma_j \psi (\beta_j \C \lambda_1) \right], \\
		p(\boldsymbol{\gamma} \C \theta) = & \displaystyle \prod_{j=1}^{p} \left[ \theta^{\gamma_j} (1-\theta)^{1-\gamma_j} \right], \\
		\theta \sim &  \mathcal{B}eta(a,b),
	\end{array}
\end{equation}
where $\psi ( \beta \C \lambda ) = (\lambda / 2) e^{-\lambda \lvert \beta \rvert}$ denotes the Laplace density with scale parameter $\lambda$.   Figure \ref{fig:laplace} depicts the Laplace density for two different choices of scale parameter. We see that for large $\lambda$ ($\lambda=20$), the density is very peaked around zero, while for small $\lambda$ ($\lambda = 1$), it is diffuse. Therefore, in our prior \eqref{SSLprior}, we typically set $\lambda_0 \gg \lambda_1$, so that $\psi ( \cdot \C \lambda_0)$ is the ``spike'' and $\psi ( \cdot\C\lambda_1 )$ is the ``slab.''

The original SSL model of \cite{RockovaGeorge2018} assumed known variance $\sigma^2=1$. \cite{MoranRockovaGeorge2018} extended the SSL to the unknown variance case. As $\sigma^2$ is typically unknown, we consider the hierarchical formulation in \cite{MoranRockovaGeorge2018} in this paper and place an independent Jeffreys prior on $\sigma^2$,
\begin{align*}
	p(\sigma^2) \propto \sigma^{-2}.
\end{align*}
Note that unlike the mixture of normals \eqref{spikeandslabnormals}, we do \textit{not} scale the Laplace priors in $p( \boldsymbol{\beta} \C \gamma)$ by $\sigma^2$. \cite{MoranRockovaGeorge2018} showed that such scaling severely underestimates the variance $\sigma^2$ when $\boldsymbol{\beta}$ is sparse or when $p > n$, thus making the model prone to overfitting.

By choosing $\lambda_1 = \lambda_0$ in \eqref{SSLprior}, we obtain the familiar LASSO $\ell_1$ penalty. On the other hand, if $\lambda_0 \rightarrow \infty$, we obtain the ``theoretically ideal'' point-mass spike-and-slab \eqref{pointmassspikeandslab} as a limiting case. Thus, a feature of the SSL prior is its ability to induce a nonconcave continuum between the penalized likelihood and (point-mass) spike-and-slab constructs.

Since it is a mixture of two Laplace distributions, the SSL prior \eqref{SSLprior} can be seen as a two-group refinement of the LASSO's $\ell_1$ penalty on the coefficients. Thus, the posterior mode for $p(\boldsymbol{\beta} \C \boldsymbol{y})$ under \eqref{SSLprior} is \textit{exactly} sparse and can be used to perform simultaneous variable selection and parameter estimation. This automatic modal thresholding property offers an advantage over previous spike-and-slab formulations \eqref{pointmassspikeandslab}-\eqref{spikeandslabnormals} which do not give exactly sparse estimates of the coefficients and which typically require \textit{post hoc} thresholding of the posterior inclusion probabilities for selection. 

It is well-known that the original LASSO \cite{Tibshirani1996} suffers from estimation bias, wherein coefficients with large magnitude are overshrunk. One may wonder what advantages the SSL \eqref{SSLprior} confers over penalized likelihood approaches such as the adaptive LASSO, SCAD, or MCP penalties \cite{FanLi2001, Zhang2010,Zou2006} which are designed to mitigate the bias problem of the LASSO. In what follows, we discuss two major advantages of the SSL. First, we demonstrate that the SSL mixes two LASSO ``bias'' terms \textit{adaptively} in such a way that either a very large amount of shrinkage is applied if $\lvert \beta_j \rvert$ is small or a very small amount of shrinkage is applied if $\lvert \beta_j \rvert$ is large. This is in contrast to the adaptive LASSO \cite{Zou2006} and similar penalties which assign \textit{fixed} coefficient-specific penalties and thus do not gear the coefficient-specific shrinkage towards these extremes. Second, the prior on $\theta$ in \eqref{SSLprior} ultimately renders the coordinates in $\boldsymbol{\beta}$ \textit{dependent} in the marginal prior $p(\boldsymbol{\beta})$ and the SSL penalty \textit{non}-separable. This provides the SSL with the additional ability to borrow information across coordinates and adapt to ensemble information about sparsity. 

\subsection{Selective shrinkage and self-adaptivity to sparsity}

As noted in Section \ref{PenalizedLikelihood}, any sparsity-inducing Bayesian prior can be recast {  in the penalized likelihood framework by treating} the logarithm of the marginal prior $\log p(\boldsymbol{\beta})$ as a penalty function. The SSL penalty is defined as
\begin{equation} \label{SSLpenalty}
	\textrm{pen} (\boldsymbol{\beta} ) = \log \left[ \frac{p(\boldsymbol{\beta})}{p(\boldsymbol{0}_p )} \right],
\end{equation}
where the penalty has been centered at $\boldsymbol{0}_p$, the $p$-dimensional zero vector, so that $\textrm{pen}({\boldsymbol{0}_p})= 0$ \cite{RockovaGeorge2018}. Using \eqref{SSLpenalty} and some algebra, the log posterior under the SSL prior (up to an additive constant) can be shown to be
\begin{equation} \label{logposteriorSSL}
	L(\boldsymbol{\beta}, \sigma^2) = -\frac{1}{2 \sigma^2} \lVert \boldsymbol{y} - \boldsymbol{X} \boldsymbol{\beta} \rVert_2^2 - (n+2) \log \sigma + \sum_{j=1}^{p} \textrm{pen}(\beta_j \C \theta_j),
\end{equation}
where for $j=1, \ldots, p$,
\begin{equation} \label{singletonpenalty}
	\textrm{pen}(\beta_j \C \theta_j) = - \lambda_1 \lvert \beta_j \rvert + \log[ p_{\theta_j}^{\star} (0 ) / p^{\star}_{\theta_j} (\beta_j ) ],
\end{equation}
with
\begin{equation} \label{conditionalinclusionprob}
	p^{\star}_{\theta_j} ( \beta_j ) = \frac{\theta_j \psi  (\beta_j \C \lambda_1)}{\theta_j \psi (\beta_j \C \lambda_1) + (1-\theta_j) \psi (\beta_j \C \lambda_0)},
\end{equation}
and
\begin{equation} \label{thetaj}
	\theta_j = E[\theta \C \boldsymbol{\beta}_{\setminus j}] = \int \theta p (\theta \C \boldsymbol{\beta}_{\setminus j}) d \theta.
\end{equation}
When $p$ is large, \cite{RockovaGeorge2018} noted that $\theta_j$ is very similar to $E [\theta\C \boldsymbol{\beta}] = \int \theta p (\theta \C \boldsymbol{\beta}) d \theta$ for every $j = 1, \ldots, p$. Thus, for practical purposes, we replace the individual $\theta_j$'s in \eqref{logposteriorSSL}-\eqref{thetaj} with a single $\widehat{\theta} = E [ \theta \C \boldsymbol{\beta}] $ going forward. 

The connection between the SSL and penalized likelihood methods is made clearer when considering the derivative of each singleton penalty $\textrm{pen}(\beta_j \C \widehat{\theta })$ in \eqref{singletonpenalty}. This derivative corresponds to an implicit bias term \cite{RockovaGeorge2018} and is given by
\begin{equation} \label{penaltyderivative}
	\frac{ \partial \textrm{pen} ( \beta_j \C \widehat{\theta} )}{\partial \lvert \beta_j \rvert } = - \lambda^{\star}_{\hat{\theta}} ( \beta_j ),
\end{equation}
where
\begin{equation} \label{lambdastar}
	\lambda^{\star}_{\hat{\theta}} (\beta_j) = \lambda_1 p^{\star}_{\hat{\theta}} ( \beta_j ) + \lambda_0 [1 - p^{\star}_{\hat{\theta}} (\beta_j)].
\end{equation}
The Karush-Kuhn-Tucker (KKT) conditions yield the following necessary condition for the global mode $\widehat{\boldsymbol{\beta}}$:
\begin{equation} \label{necessarycond}
	\widehat{\beta}_j = \frac{1}{n} \left[ \lvert z_j \rvert - \sigma^2 \lambda^{\star}_{\hat{\theta}} (\widehat{\beta}_j) \right]_{+} \textrm{sign}(z_j), \hspace{.5cm} j=1, \ldots, p,
\end{equation}
where $z_j = \boldsymbol{x}_j^T (\boldsymbol{y} - \sum_{k \neq j} \boldsymbol{x}_k \widehat{\beta}_k )$. Notice that the condition \eqref{necessarycond} resembles the soft-thresholding operator for the LASSO, except that it contains an adaptive penalty term $\lambda_{\hat{\theta}}^{\star}$ for \textit{each} coefficient. In particular, the quantity \eqref{lambdastar} is a weighted average of the two regularization parameters, $\lambda_1$ and $\lambda_0$, and the weight $p_{\hat{\theta}}^{\star} (\beta_j)$. Thus, \eqref{lambdastar}-\eqref{necessarycond} show that the SSL penalty induces an \textit{adaptive} regularization parameter which applies a different amount of shrinkage to each coefficient, unlike the original LASSO which applies the same shrinkage to every coefficient. 

It is worth looking at the term $p_{\hat{\theta}}^{\star} (\beta_j)$ more closely. In light of \eqref{conditionalinclusionprob}, this quantity can be viewed as a conditional probability that $\beta_j$ was drawn from the slab distribution rather than the spike distribution, having seen the regression coefficient $\beta_j$. We have $p_{\hat{\theta}}^{\star} (\beta_j) = \Pr (\gamma_j = 1 \C \beta_j, \hat{\theta})$, where
\begin{equation} \label{pstarexpanded}
	p_{\hat{\theta}}^{\star} (\beta_j) = \frac{1}{1 + \frac{(1-\hat{\theta})}{\hat{\theta}} \frac{\lambda_0}{\lambda_1} \exp \left[ - \lvert \beta_j \rvert ( \lambda_0 - \lambda_1 ) \right]}
\end{equation}
is an \textit{exponentially increasing} function in $\lvert \beta_j \rvert$. From \eqref{pstarexpanded}, we see that the functional $p_{\hat{\theta}}^{\star}$ has a sudden increase from near-zero to near-one. Therefore, $p_{\hat{\theta}}^{\star} (\beta_j)$ gears $\lambda_{\hat{\theta}}^{\star}$ in \eqref{lambdastar} towards the extreme values $\lambda_1$ and $\lambda_0$, depending on the size of $\lvert \beta_j \rvert$. Assuming that $\lambda_1$ is sufficiently small (and hence, the slab $\psi (\beta_j \C \lambda_1)$ is sufficiently diffuse), this allows the large coefficients to escape the overall shrinkage effect, in sharp contrast to the single Laplace distribution \eqref{bayesianLASSO}, where the bias issue remains even if a prior is placed on $\lambda$ \cite{RockovaGeorge2016Abel}.

Apart from its selective shrinkage property, a second key benefit of the SSL model \eqref{SSLprior} is its \textit{self-adaptivity} to the sparsity pattern of the data through the prior on the mixing proportion $\theta$, $p(\theta) \sim \mathcal{B}eta(a,b)$. As mentioned previously, this prior ultimately renders the SSL penalty \textit{non}-separable. Fully separable penalty functions, such as those described in Section \ref{PenalizedLikelihood}, are limited by their inability to adapt to common features across model parameters because they treat these parameters independently. In contrast, treating $\theta$ (the expected proportion for non-negligible coefficients in $\boldsymbol{\beta}$) as random, allows for automatic adaptivity to different levels of sparsity. As shown in \eqref{conditionalinclusionprob}-\eqref{thetaj} (and replacing $\theta_j$ with $\hat{\theta}$ and $\boldsymbol{\beta}_{\setminus j}$ with $\boldsymbol{\beta}$), the mixing weight $p^{\star}_{\hat{\theta}}$ is obtained by averaging $p_{\theta}^{\star} (\cdot)$ over $p(\theta \C \boldsymbol{\beta})$, i.e. $p_{\hat{\theta}}^{\star} (\beta) = \int_{0}^{1} p_{\theta}^{\star} (\beta_j) p(\theta \C \boldsymbol{\beta}) d \theta$. It is through this averaging that the SSL penalty \eqref{SSLpenalty} is given an opportunity to borrow information across coordinates and learn about the underlying level of sparsity in $\boldsymbol{\beta}$. 

For the hyperparameters $(a,b)$ in the beta prior on $\theta$, \cite{RockovaGeorge2018} recommended the default choice of $a=1, b=p$. By Lemma 4 of \cite{RockovaGeorge2018}, this choice ensures that $E[\theta \C \widehat{\boldsymbol{\beta}}] \sim \widehat{p}_{\gamma}/p$, where $\widehat{p}_{\gamma}$ is the number of nonzero coefficients in $\boldsymbol{\widehat{\beta}}$. Further, this choice of hyperparameters results in an automatic multiplicity adjustment \cite{ScottBerger2010} and ensures that $\theta$ is small (or that most of the coefficients belong to the spike) with high probability. Thus, the SSL also favors parsimonious models in high dimensions and avoids the curse of dimensionality.

\subsection{The spike-and-slab LASSO in action}

Before delving into the implementation details of the SSL, we perform a small simulation study to illustrate the benefits of the adaptive shrinkage of SSL versus the non-adaptive shrinkage of the LASSO. We simulated data of $n=50$ observations with $p=12$ predictors generated as four independent blocks of highly correlated predictors. More precisely, $n$ rows of our design matrix $\boldsymbol{X}$ were generated independently from a $\mathcal{N}_p (\boldsymbol{0}, \boldsymbol{\Sigma})$ distribution with block diagonal covariance matrix $\boldsymbol{\Sigma} = \textrm{bdiag} ( \widetilde{\boldsymbol{\Sigma}}, \ldots, \widetilde{\boldsymbol{\Sigma}})$, where $\widetilde{\boldsymbol{\Sigma}} = \{\widetilde{\sigma}_{ij} \}_{i,j=1}^{3}, \widetilde{\sigma}_{ij} = 0.9$  if $i \neq j$ and $\widetilde{\sigma}_{ii} = 1$. The response was generated from $\boldsymbol{y} \sim \mathcal{N}_n ( \boldsymbol{X} \boldsymbol{\beta}_0, \boldsymbol{I} )$, with $\boldsymbol{\beta}_0 = (1.3, 0, 0, 1.3, 0, 0, 1.3, 0, 0, 1.3, 0, 0)'$. Note that only $x_1$, $x_4$, $x_7$, and $x_{10}$ are non-null in this true model.

\begin{figure}[t!]
	\centering
	\includegraphics[width=0.49\textwidth]{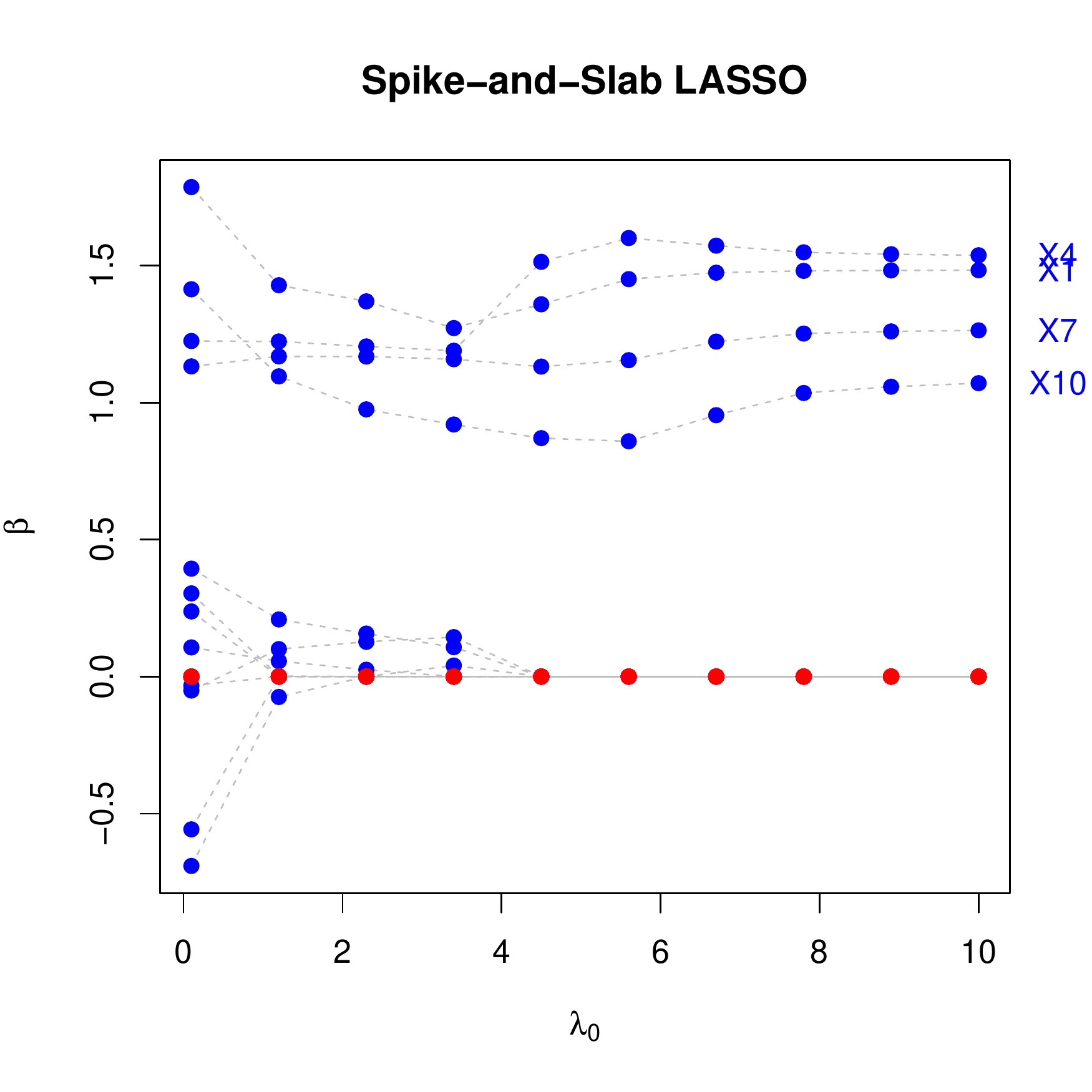}
	\includegraphics[width=0.49\textwidth]{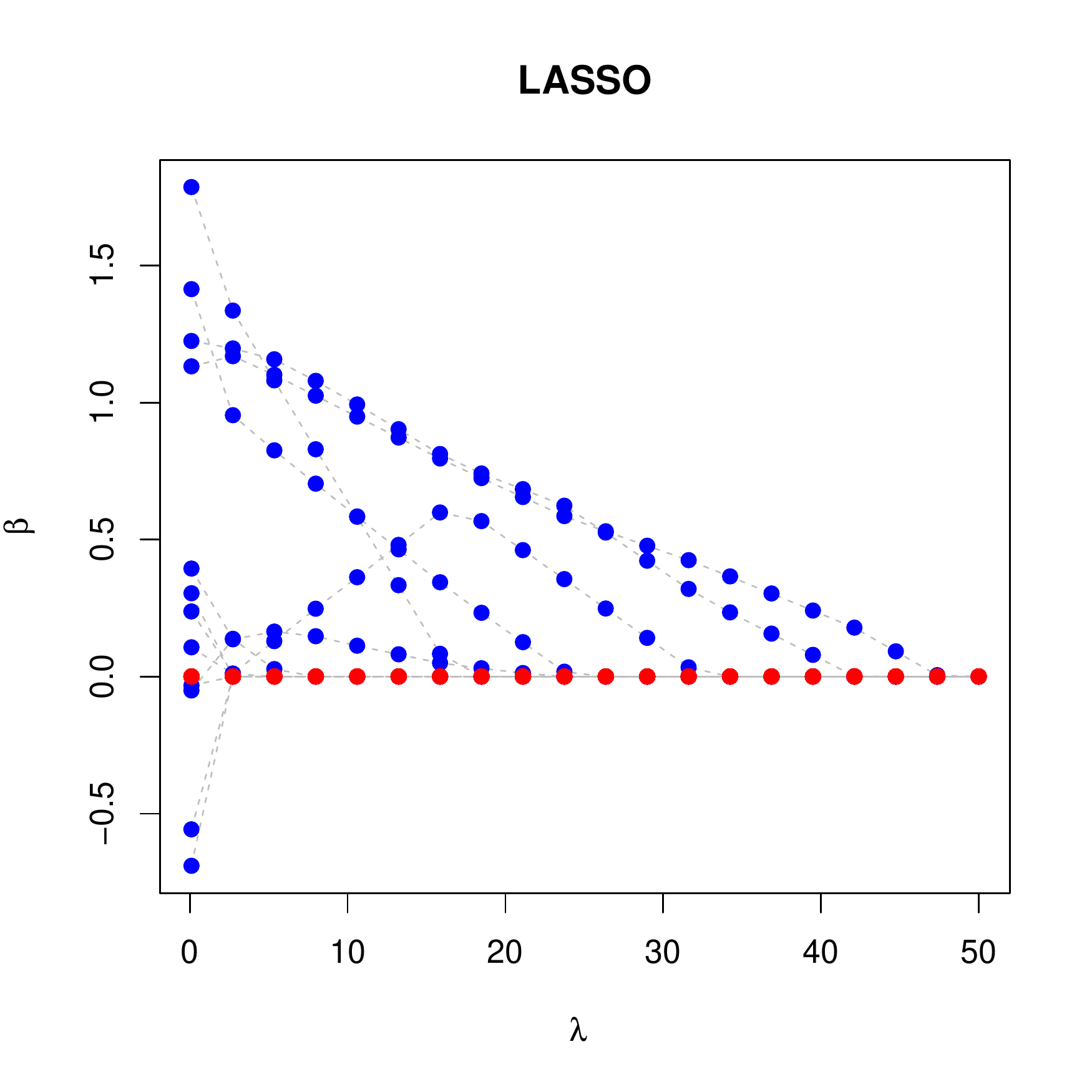}
	\caption{The coefficient paths of $\widehat{\boldsymbol{\beta}}^{\textrm{SSL}}$ (left panel) and $\widehat{\boldsymbol{\beta}}^{\textrm{LASSO}}$ (right panel) as $\lambda_0$ and $\lambda$ respectively are increased. The connected points off the horizontal axis are the nonzero estimates. The points along the horizontal axis are the zero values where the negligible estimates disappear.}
	\label{fig:coefficientpaths}
\end{figure}

We fit both the SSL and the LASSO of \cite{Tibshirani1996} to this model. Figure \ref{fig:coefficientpaths} displays the coefficient paths for both SSL and LASSO as the spike parameter $\lambda_0$ in the SSL and the regularization parameter $\lambda$ in the LASSO are increased. For the SSL, the spike parameter $\lambda_1 = 0.01$ is fixed throughout. Both the SSL and LASSO begin at $\lambda_0 = \lambda = 0$ with the same 12 (nonzero) ordinary least squares (OLS) estimates for $\boldsymbol{\beta}_0$. However, as $\lambda_0$ increases for the SSL,  the eight smaller OLS estimates are gradually shrunk to zero by the SSL's spike. Meanwhile, the four large estimates are held steady by the SSL's slab, eventually stabilizing at values close to their OLS estimates. The SSL correctly selects the four nonzero coefficients in the true model, demonstrating its self-adaptivity to the true sparsity pattern of the data.

In contrast, Figure \ref{fig:coefficientpaths} also shows that as the LASSO's single penalty parameter $\lambda$ increases, \textit{all} twelve estimates are gradually shrunk to zero. This is because without a slab distribution to help hold the large values steady, the LASSO eventually shrinks all estimates to zero for a large enough $\lambda$. Additionally, due to the order in which the 12 estimates have been thresholded to zero, no value of $\lambda$ yields the correct subset selection $\{ x_1, x_4, x_y, x_{10} \}$. In particular, if $\lambda$ is chosen from cross-validation, the LASSO selects a subset of variables with four false positives.   

Our small simulation study illustrates the advantage of the SSL over the LASSO. Specifically, because the LASSO applies the same amount of shrinkage to all regression coefficients, it may estimate a null model if its regularization parameter $\lambda$ is too large. The SSL's two-group refinement of the LASSO penalty helps to mitigate this problem by facilitating selective shrinkage. In Section \ref{Illustration}, we further illustrate the strong empirical performance of the SSL in high-dimensional settings when $p > n$.

\section{Computational details} \label{Computing}

We now turn our attention to implementation of the SSL model \eqref{SSLprior} under the normal linear regression model \eqref{linearregression}. The method described in Section \ref{MAPAlgorithm} is implemented in the publicly available \textsf{R} package \texttt{SSLASSO} \cite{SSLASSOpackage}. However, we also describe an alternative implementation approach in Section \ref{EMAlgorithm}, which is amenable to situations outside of the Gaussian likelihood.

\subsection{Coordinate-wise optimization} \label{MAPAlgorithm}

As mentioned in Section \ref{SpikeAndSlabLASSO} and shown in \eqref{necessarycond}, the (global) posterior mode under the SSL prior \eqref{SSLprior} is exactly sparse, while avoiding the excessive bias issue for large coefficients. Therefore, we can obtain estimates for $\boldsymbol{\beta}$ by targeting the posterior mode. 

Marginalizing out $\boldsymbol{\gamma}$ in \eqref{SSLprior} gives the prior for $\boldsymbol{\beta}$ (conditional on $\theta$),
\begin{equation} \label{SSLconditionalontheta}
	p(\boldsymbol{\beta} \C \theta) = \prod_{j=1}^{p} \left[ (1-\theta) \psi (\beta_j \C \lambda_0) + \theta \psi (\beta_j \C \lambda_1) \right].
\end{equation}
Using this reparametrization, \cite{MoranRockovaGeorge2018,RockovaGeorge2018} developed a highly efficient coordinate ascent algorithm to quickly target the mode for $(\boldsymbol{\beta}, \sigma^2)$.

Since the SSL is a non-convex method, the KKT conditions only give a necessary condition \eqref{necessarycond} for $\widehat{\boldsymbol{\beta}}$ to be a global mode, but not a sufficient one. When $p > n$ and $\lambda_0 \gg \lambda_1$, the posterior will typically be multimodal. Nevertheless, it is still possible to obtain a refined characterization of the global mode. Building upon theory developed by \cite{ZhangZhang2012}, \cite{MoranRockovaGeorge2018,RockovaGeorge2018} gave necessary \textit{and} sufficient conditions for $\widehat{\boldsymbol{\beta}}$ to be a \textit{global} mode. By Theorems 3-4 in \cite{RockovaGeorge2018} and Propositions 4-5 in \cite{MoranRockovaGeorge2018}, the global mode under the SSL prior \eqref{SSLprior} is a blend of soft-thresholding \textit{and} hard-thresholding, namely
\begin{equation} \label{globalmode}
	\widehat{\beta}_j = \frac{1}{n} \left[ | z_j | - \sigma^2 \lambda_{\hat{\theta}}^{\star} (\widehat{\beta}_j) \right]_{+} \textrm{sign}(z_j) \mathbb{I}( |z_j| > \Delta ),
\end{equation}
where $z_j = \boldsymbol{x}_j^T (\boldsymbol{y} - \sum_{k \neq j} \boldsymbol{x}_k \widehat{\beta}_k)$ and $\Delta \equiv \inf_{t >0} [nt/2 - \sigma^2 \textrm{pen}(t \C\hat{\theta} ) / t]$. In \cite{MoranRockovaGeorge2018}, an approximation for $\Delta$ is given by
\begin{align*}
	\Delta = \left\{ 
	\begin{array}{ll}
		\sqrt{2 n \sigma^2 \log [ 1 / p_{\hat{\theta}}^{\star} (0)]} + \sigma^2 \lambda_1\ & \textrm{if } g_{\hat{\theta}}(0) > 0, \\
		\sigma^2 \lambda_{\hat{\theta}}^{\star} (0) & \textrm{otherwise},
	\end{array}
	\right.
\end{align*}
where $g_{\theta}(x) = [ \lambda_{\theta}^{\star}(x) - \lambda_1]^2 + (2n / \sigma^2) \log p_{\theta}^{\star} (x)$. The generalized thresholding operator \eqref{globalmode} allows us to eliminate many suboptimal local modes from consideration through the threshold $\Delta$. This refined characterization also facilitates a highly efficient coordinate ascent algorithm \cite{MazumderFriedmanHastie2011} to find the global mode, which we now detail.

After initializing $(\Delta^{(0)}, \boldsymbol{\beta}^{(0)}, \theta^{(0)}, \sigma^{2(0)})$, the coordinate ascent algorithm iteratively updates these parameters until convergence. The update for the threshold $\Delta$ at the $t^{th}$ iteration is 
\begin{align*}
	\Delta^{(t)} = \left\{ 
	\begin{array}{ll}
		\sqrt{2 n \sigma^{2(t-1)} \log[1 / p_{\hat{\theta}^{(t-1)}}^{\star}(0)]} + \sigma^{2(t-1)} \lambda_1 & \textrm{if } g_{\theta^{(t-1)}} (0) > 0, \\
		\sigma^{2(t-1)} \lambda_{\theta^{(t-1)}}^{\star}(0) & \textrm{otherwise}.
	\end{array}
	\right.
\end{align*}
Next, $\boldsymbol{\beta}$ is updated as 
\begin{align*}
	\beta_j^{(t)} \leftarrow \frac{1}{n} \left( |z_j| - \lambda_{\hat{\theta}^{(t-1)}}^{\star} ( \widehat{\beta}_j^{(t-1)} ) \right)_{+} \textrm{sign}(z_j) \mathbb{I}( |z_j| > \Delta^{(t)}).
\end{align*}
Using the approximation for $E[\theta \C \widehat{\boldsymbol{\beta}}]$ in Lemma 4 of \cite{RockovaGeorge2018}, the update for $\hat{\theta}$ is
\begin{align*}
	\hat{\theta}^{(t)} \leftarrow \frac{a + \widehat{p}_{\gamma}^{(t)}}{a+b+p},
\end{align*}
where $\widehat{p}_{\gamma}^{(t)}$ is the number of nonzero entries in $\boldsymbol{\beta}^{(t)}$. Finally, the update for $\sigma^2$ is
\begin{align*}
	\sigma^{2(t)} \leftarrow \frac{ \lVert \boldsymbol{y} - \boldsymbol{X} \boldsymbol{\beta}^{(t)} \rVert_2^2}{n+2}.
\end{align*}

\subsection{Dynamic posterior exploration} \label{DynamicPosteriorExploration}

The performance of the SSL model depends on good choices for the hyperparameters $(\lambda_0, \lambda_1)$ in \eqref{SSLprior}. To this end, \cite{RockovaGeorge2018} recommend a ``dynamic posterior exploration'' strategy in which the slab hyperparameter $\lambda_1$ is held fixed at a small value and the spike hyperparameter $\lambda_0$ is gradually increased along a ladder of increasing values, $\{ \lambda_0^{1}, \ldots, \lambda_0^{L} \}$. The algorithm is not very sensitive to the specific choice of $\lambda_1$, provided that the slab is sufficiently diffuse. For each $\lambda_0^{s}$ in the ladder for the spike parameters, we reinitialize $(\Delta^{(0)}, \boldsymbol{\beta}^{(0)}, \theta^{(0)}, \sigma^{2(0)})$ using the MAP estimates for these parameters from the previous spike parameter $\lambda_0^{s-1}$ as a ``warm start.'' 

This sequential reinitialization strategy allows the SSL to more easily find the global mode. In particular, when $(\lambda_1 - \lambda_0)^2 < 4$ and $\sigma^2$ is fixed, the objective \eqref{logposteriorSSL} is convex. The intuition here is to use the solution to the convex problem as a ``warm'' start for the non-convex problem (when $\lambda_0 \gg \lambda_1$). As we increase $\lambda_0$, the posterior becomes ``spikier,'' with the spikes absorbing more and more of the negligible parameters. Meanwhile, keeping $\lambda_1$ fixed at a small value allows the larger coefficients to escape the pull of the spike. For large enough $\lambda_0$, the algorithm will eventually stabilize so that further increases in $\lambda_0$ do not change the solution. In Section \ref{Illustration}, we illustrate this with plots of the SSL solution paths.

Additionally, as noted by \cite{MoranRockovaGeorge2018}, some care must also be taken when updating $\sigma^2$. When $p > n$ and $\lambda_0 \approx \lambda_1$, the model can become saturated, causing the residual variance to go to zero. To avoid this suboptimal mode at $\sigma^2 = 0$, \cite{MoranRockovaGeorge2018} recommend fixing $\sigma^2$ until the $\lambda_0$ value in the ladder at which the algorithm starts to converge in less than 100 iterations. Then, $\boldsymbol{\beta}$ and $\sigma^2$ are simultaneously updated for the next largest $\lambda_0$ in the sequence. The complete algorithm for coordinate-wise optimization with dynamic posterior exploration is given in Section 4 of the supplementary material in \cite{MoranRockovaGeorge2018}. This algorithm is implemented in the \textsf{R} package \texttt{SSLASSO}.

\subsection{EM implementation of the spike-and-slab LASSO} \label{EMAlgorithm}



The coordinate ascent algorithm of Section \ref{MAPAlgorithm} specifically appeals to the theoretical framework of \cite{ZhangZhang2012} to search for the global SSL mode $\boldsymbol{\widehat{\beta}}$.  An alternative approach, also proposed by  \cite{RockovaGeorge2018},  is to use an EM algorithm in the vein of EMVS \cite{RockovaGeorge2014}.    Again treating the latent variables $\boldsymbol{\gamma}$ in \eqref{SSLprior} are treated as ``missing'' data, this EM implementation of the SSL proceeds as follows.

In the E-step at the $t$th iteration, we compute $E[\tau_j \C \boldsymbol{y}, \boldsymbol{\beta}^{(t-1)}, \theta^{(t-1)}, \sigma^{2(t-1)}] =  p_{\theta^{(t-1)}}^{\star} (\beta_j^{(t-1)})$, where $p_{\theta}^{\star}$ is as in \eqref{conditionalinclusionprob}. The M-step then iterates through the following updates:
\begin{align*}
	& \boldsymbol{\beta}^{(t)} \leftarrow \displaystyle \argmax_{\boldsymbol{\beta} \in \mathbb{R}^{p}} \left\{ - \frac{1}{2} \lVert \boldsymbol{y} - \boldsymbol{X} \boldsymbol{\beta} \rVert_2^2 - \sum_{j=1}^{p} \sigma^{2(t-1)} \lambda_{\theta^{(t-1)}}^{\star}(\beta_j^{(t-1)}) \lvert \beta_j \rvert \right\}, \\
	& \theta^{(t)} \leftarrow \frac{ \sum_{j=1}^{p} p_{\theta^{(t-1)}}^{\star} (\beta_j^{(t)}) + a - 1}{a+b+p - 2}, 
	\\
	& \sigma^{2(t)} \leftarrow \frac{ \lVert \boldsymbol{y} - \boldsymbol{X} \boldsymbol{\beta}^{(t)} \rVert_2^2}{n+2},
\end{align*}
where $\lambda_{\theta}^{\star} (\beta) = \lambda_1 p_{\theta}^{\star} (\beta) + \lambda_0 [ 1- p_{\theta}^{\star} (\beta) ]$. Note that the update for $\boldsymbol{\beta}^{(t)}$ is an adaptive LASSO regression with weights $\sigma^2 \lambda_{\theta}^{\star}$ and hence can be solved very efficiently using coordinate descent algorithms \cite{FriedmanHastieTibshirani2010}. Like EMVS \cite{RockovaGeorge2014}, the dynamic posterior exploration strategy detailed in Section \ref{DynamicPosteriorExploration} can be used to find a more optimal mode for $(\boldsymbol{\beta}, \sigma^2)$.

This EM approach can be straightforwardly adapted for other statistical models where the SSL prior \eqref{SSLprior} is used (such as the methods described in Section \ref{MethodologicalExtensions}) but where the likelihood function differs and the theory of \cite{ZhangZhang2012} is not applicable. 
Similar to the coordinate ascent algorithm described in Section \ref{MAPAlgorithm}, this EM algorithm may be sensitive to the initialization of $(\boldsymbol{\beta}^{(0)}, \theta^{(0)}, \sigma^{2(0)} )$. The dynamic posterior exploration strategy described earlier can partly help to mitigate this issue, since the posterior starts out relatively flat when $\lambda_0 \approx \lambda_1$ but becomes ``spikier'' as $\lambda_0$ increases. By the time that the spikes have reappeared, the ``warm start'' solution from the previous $\lambda_0$ in the ladder should hopefully be in the basin of dominant mode. Other strategies such as running the algorithm for a wide choice of starting values or deterministic annealing can also aid in adding robustness against poor initializations \cite{McLachlanBasford1988,RockovaGeorge2014,UedaNakano1998}.

\section{Uncertainty quantification} \label{UncertaintyQuantification}

While the algorithms described in Section \ref{Computing} can be used to rapidly target the modes of the SSL posterior, providing a measure of uncertainty for our estimates is a challenging task. In this section, we outline two possible strategies for the task of uncertainty quantification. The first is based on debiasing the posterior mode. The second involves posterior simulation. 

\subsection{Debiasing the posterior mode}

One possible avenue for uncertainty quantification is to use debiasing \cite{BaiMoranAntonelliChenBoland2019, JavanmardMontanari2018, VanDeGeerBuhlmannRitovDezeure2014,ZhangZhang2014}. Let $\widehat{\boldsymbol{\Sigma}} = \boldsymbol{X}^T \boldsymbol{X} / n$ and let $\widehat{\boldsymbol{\Theta}}$ be an approximate inverse of $\widehat{\boldsymbol{\Sigma}}$. Note that when $p > n$, $\boldsymbol{X}$ is singular, so $\widehat{\boldsymbol{\Sigma}}^{-1}$ does not necessarily exist. However, we can still obtain a sparse estimate of the precision matrix $\widehat{\boldsymbol{\Theta}}$ for the rows of $\boldsymbol{X}$ by using techniques from the graphical models literature, e.g. the nodewise regression procedure in \cite{MeinshausenBuhlmann2006} or the graphical lasso \cite{FriedmanHastieTibshirani2007}. We define the quantity $\widehat{\boldsymbol{\beta}}_d$ as
\begin{equation}
	\widehat{\boldsymbol{\beta}}_d = \widehat{\boldsymbol{\beta}} + \widehat{\boldsymbol{\Theta}} \boldsymbol{X}^T (\boldsymbol{y} - \boldsymbol{X} \widehat{\boldsymbol{\beta}})/n.
\end{equation}
where $\widehat{\boldsymbol{\beta}}$ is the MAP estimator of $\boldsymbol{\beta}$ under the SSL model. By \cite{VanDeGeerBuhlmannRitovDezeure2014}, this quantity $\widehat{\boldsymbol{\beta}}_d$ has the following asymptotic distribution: 
\begin{equation} \label{asymptoticdist}
	\sqrt{n}(\widehat{\boldsymbol{\beta}}_d - \boldsymbol{\beta}) \sim \mathcal{N}(\boldsymbol{0}, \sigma^2 \widehat{\boldsymbol{\Theta}} \widehat{\boldsymbol{\Sigma}} \widehat{\boldsymbol{\Theta}}^T).
\end{equation}
For inference, we replace the population variance $\sigma^2$ in \eqref{asymptoticdist} with the modal estimate $\widehat{\sigma}^2$ from the SSL model. Let $\widehat{\beta}_{dj}$ denote the $j$th coordinate of $\widehat{\boldsymbol{\beta}}_d$. We have from \eqref{asymptoticdist} that the $100(1-\alpha)  \%$ asymptotic pointwise confidence intervals for $\beta_{j}, j = 1, \ldots, p$, are
\begin{align} \label{confidenceintervals}
	[ \widehat{\beta}_{dj} - c(\alpha, n, \widehat{\sigma}^2), \widehat{\beta}_{dj} + c(\alpha, n, \widehat{\sigma}^2) ],
\end{align}
where $c(\alpha, n, \widehat{\sigma}^2) := \Phi^{-1} (1-\alpha/2) \sqrt{ \widehat{\sigma}^2 ( \widehat{\boldsymbol{\Theta}} \widehat{\boldsymbol{\Sigma}} \widehat{\boldsymbol{\Theta}}^T )_{jj} / n}$ and $\Phi(\cdot)$ denotes the cumulative distribution function of $\mathcal{N}(0,1)$.

Note that the posterior modal estimate $\widehat{\boldsymbol{\beta}}$ under the SSL prior already has much less bias than the LASSO estimator \cite{Tibshirani1996}. Therefore, the purpose of the debiasing procedure above is mainly to obtain an estimator with an asymptotically normal distribution from which we can construct asymptotic pointwise confidence intervals. While this procedure is asymptotically valid, \cite{AntonelliParmigianiDominici2019} showed through numerical studies that constructing confidence intervals based on asymptotic arguments may provide coverage below the nominal level in finite samples, especially small samples. Therefore, it may be more ideal to use the actual SSL posterior $p(\boldsymbol{\beta} \C \boldsymbol{y})$ for inference. 

\subsection{Posterior sampling for the spike-and-slab LASSO}
Fully Bayesian inference with the SSL can be carried out via posterior simulation.
However, posterior sampling under spike-and-slab priors has continued to pose challenges.  
One immediate strategy for sampling from the SSL posterior is the SSVS algorithm of \cite{GeorgeMcCulloch1993}, described in Section \ref{sspriors}. One can regard the Laplace distribution  as a scale mixture of Gaussians with an exponential mixing distribution \cite{ParkCasella2008} and perform a variant of SSVS. Recently, several clever computational tricks have been suggested that avoid costly matrix inversions needed by SSVS by using linear solvers \cite{BhattacharyaChakrabortyMallick2016}, low-rank approximations \cite{JohndrowOrensteinBhattacharya2020}, or by disregarding correlations between active and inactive coefficients \cite{NarisettyShenHe2019}. These techniques can be suitably adapted for fast posterior sampling of the SSL as well.

Intrigued by the speed of SSL mode detection, \cite{NieRockova2020} explored the possibility of turning SSL into approximate posterior sampling by performing MAP optimization on many independently perturbed datasets.  Building on Bayesian bootstrap ideas, they introduced a method for approximate sampling called Bayesian bootstrap spike-and-slab LASSO (BB-SSL) which scales linearly with both $n$ and $p$.  Beyond its scalability, they show that BB-SSL has strong theoretical support, matching the convergence rate of the original posterior in sparse normal-means and in high-dimensional regression.  

\section{Illustrations} \label{Illustration} 

In this section, we illustrate the SSL's potential for estimation, variable selection, and prediction on both simulated and real high-dimensional data sets.

\subsection{Example on synthetic data} \label{Simulations}

For our simulation study, we slightly modified the settings in \cite{MoranRockovaGeorge2018}. We set $n = 100$ and $p = 1000$ in \eqref{linearregression}. The design matrix $\boldsymbol{X}$ was generated from a multivariate Gaussian distribution with mean $\boldsymbol{0}_p$ and a block-diagonal covariance matrix $\boldsymbol{\Sigma} = \textrm{bdiag} ( \widetilde{\Sigma}, \ldots, \widetilde{\Sigma} )$, where $\widetilde{\Sigma} = \{ \widetilde{\sigma} \}_{i,j=1}^{50}$, with $\widetilde{\sigma}_{ij} = 0.9$ if $i \neq j$ and $\widetilde{\sigma}_{ii} = 1$. The true vector of regression coefficients $\boldsymbol{\beta}_0$ was constructed by assigning regression coefficients $\{-3.5, -2.5, -1.5, 1.5, 2.5, 3.5 \}$ to 6 entries located at the indices $\{ 1, 51, 101, 151, 201, 251 \}$ and setting the remaining coefficients equal to zero. Hence, there were 20 independent blocks of 50 highly correlated predictors, where the first six blocks contained only one active predictor. We then generated the response $\boldsymbol{y}$ using \eqref{linearregression}, where the error variance was set as $\sigma^2 = 3$. 

We compared the SSL with the LASSO \cite{Tibshirani1996}, SCAD \cite{FanLi2001}, and MCP \cite{Zhang2010}. The SSL method was applied using the \textsf{R} package \texttt{SSLASSO}. The competing methods were applied using the \textsf{R} package \texttt{ncvreg}. We repeated our experiment 500 times with new covariates and responses generated each time. For each experiment, we recorded the mean squared error (MSE) and mean prediction error (MPE), defined as
\begin{align*}
	\textrm{MSE} = \frac{1}{p} \lVert \widehat{\boldsymbol{\beta}} - \boldsymbol{\beta}_0 \rVert_2^2 \hspace{.3cm} \textrm{ and } \hspace{.3cm}
	\textrm{MPE} = \frac{1}{n} \lVert \boldsymbol{X} ( \widehat{\boldsymbol{\beta}} - \boldsymbol{\beta}_0 ) \rVert_2^2.
\end{align*}
We also kept track of $\widehat{p}_{\gamma}$, or the size of the model selected by each of these methods. Finally, we recorded the false discovery rate (FDR), the false negative rate (FNR), and the Matthews correlation coefficient (MCC) \cite{Matthews1975}, defined respectively as
\begin{align*}
	\textrm{FDR} & = \frac{\textrm{FP}}{\textrm{TN}+\textrm{FP}}, \hspace{.5cm} \textrm{FNR} = \frac{\textrm{FN}}{\textrm{TP}+\textrm{FN}}, \\
	\textrm{MCC} & = \frac{ \textrm{TP} \times \textrm{TN} - \textrm{FP} \times \textrm{FN}}{ \sqrt{ ( \textrm{TP} + \textrm{FP} )(\textrm{TP} + \textrm{FN})(\textrm{TN} + \textrm{FP})(\textrm{TN} + \textrm{FN})}},
\end{align*}
where TP, TN, FP, and FN denote the number of true positives, true negatives, false positives, and false negatives respectively. The MCC is a correlation coefficient between the predicted set of significant coefficients and the actual set of nonzero coefficients \cite{Matthews1975}. MCC has a range of -1 to 1,  with -1 indicating completely incorrect selection (i.e. TP=TN=0) and 1 indicating completely correct variable selection (i.e. FP=FN=0). Models with MCC closer to 1 have higher selection accuracy. \textsf{R} code to reproduce these experiments is available in the online supplementary material.

\begin{table}[t!]  
	\centering
	\begin{tabularx}{\textwidth}{@{}*7{>{\raggedright\arraybackslash}X}@{}}
		\hline
		& MSE & MPE & $\widehat{p}_{\gamma}$ & FDR & FNR & MCC \\ 
		\hline
		\hline
		SSL & \textbf{0.0067} (0.0076) & \textbf{0.701} (0.542) & \textbf{6.05} (0.271) & \textbf{0.0012} (0.0010) & 0.187 (0.160) & \textbf{0.809} (0.162) \\
		\hline
		LASSO & 0.011 (0.0045) & 1.14 (0.303) & 33.38 (5.38) & 0.028 (0.0055) & \textbf{0.083} (0.109) & 0.387 (0.062) \\
		\hline 
		SCAD & 0.011 (0.012) & 0.985 (0.691) & 12.74 (3.98) & 0.0081 (0.0043) & 0.225 (0.187) & 0.554 (0.178) \\
		\hline
		MCP & 0.020 (0.016) & 1.55 (0.849) & 11.31 (2.70) & 0.0077 (0.0031) & 0.395 (0.211) & 0.447 (0.173) \\
		\hline
	\end{tabularx}
	\caption{MPE, MPE, estimated model size, FDR, FNR, and MCC for SSL, LASSO, SCAD, and MCP.  The results are averaged across 500 replications. In parentheses, we report the empirical standard errors.}  
	\label{Table:1}
\end{table}

Table \ref{Table:1} reports our results averaged across the 500 replications. We see that the SSL had the lowest average MSE and MPE, in addition to selecting (on average) the most parsimonious model. The LASSO (along with SCAD) had the second lowest MSE, but it tended to select far more variables than the other methods, leading to the highest FDR. In contrast, SSL had the lowest FDR and the highest MCC, indicating that the SSL had the best overall variable selection performance of all the methods. Our simulation study demonstrates that SSL achieves both parsimony \textit{and} accuracy of estimation and selection. 

\begin{figure}[t!]
	\centering
	\includegraphics[width=\textwidth]{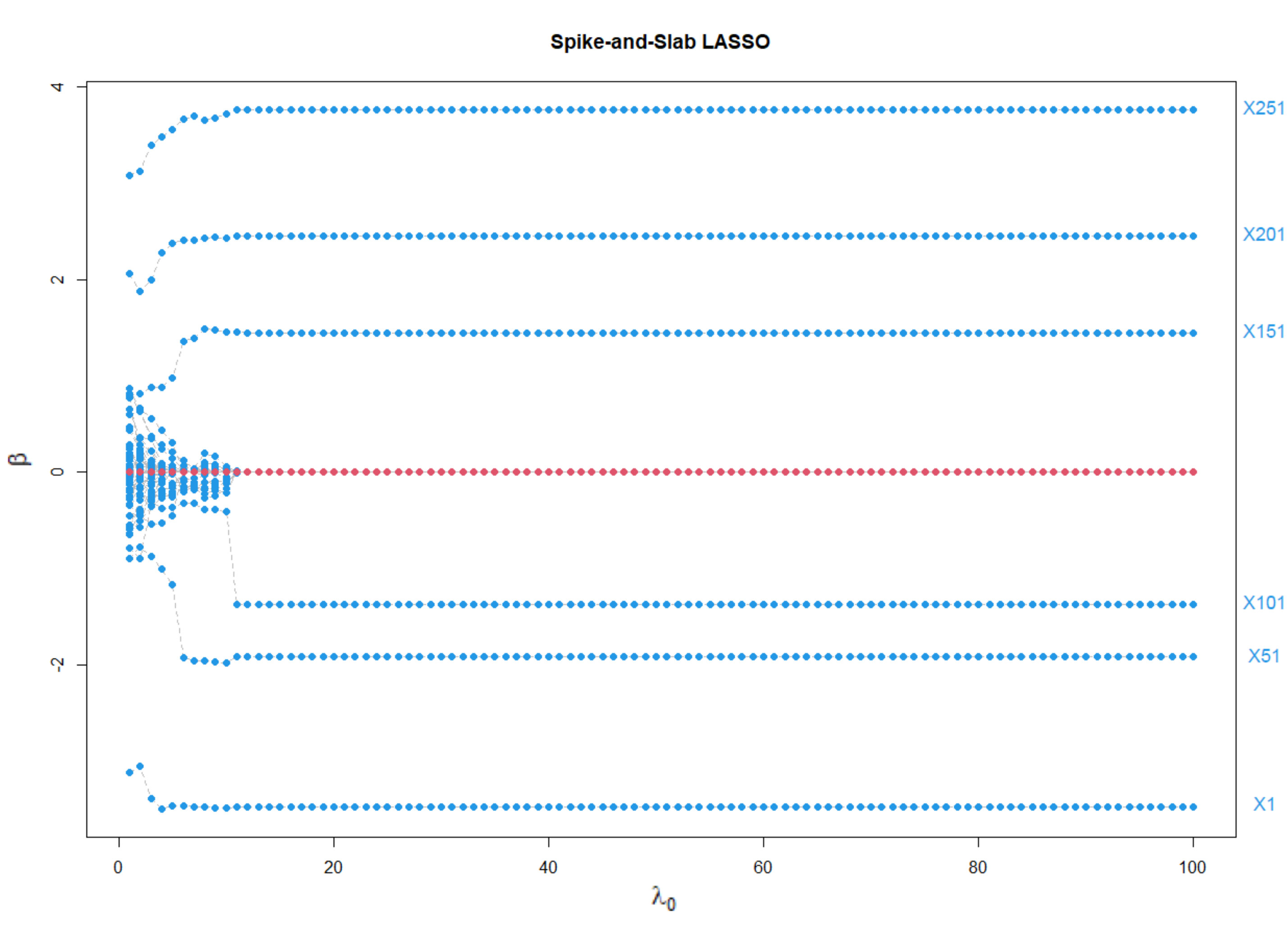} 
	\caption{The solution paths for the SSL along the ladder of spike parameters $\lambda_0$. We see that the solution stabilizes after a certain point, so that further increases in $\lambda_0$ do not change the solution.  The points along the horizontal axis are the zero values where the negligible estimates disappear. }
	\label{fig:dynamicposteriorexploration}
\end{figure}

Figure \ref{fig:dynamicposteriorexploration} illustrates the benefits of the dynamic posterior exploration approach outlined in Section \ref{DynamicPosteriorExploration}. Specifically, Figure \ref{fig:dynamicposteriorexploration} plots the solution paths for the regression coefficients from one of our experiments as the spike hyperparameter $\lambda_0$ increases. We see that the SSL solution stabilizes fairly quickly (when $\lambda_0$ is less than 20), so that further increases in $\lambda_0$ do not change the solution. This demonstrates that dynamic posterior exploration offers a viable alternative to cross-validation. The \textsf{R} package \texttt{SSLASSO} provides the functionality to generate plots of these solution paths. 

\subsection{Bardet-Beidl syndrome gene expression study} \label{RealDataAnalysis}

We now analyze a microarray data set consisting of gene expression measurements from the eye tissue of 120 laboratory rats\footnote{Data accessed from the Gene Expression Omnibus \texttt{ www.ncbi.nlm.nih.gov/geo} (accession no. GSE5680).}. The data was originally studied by \cite{Scheetz06} to investigate mammalian eye disease. In this data, the goal is to identify genes which are associated with the gene TRIM32. TRIM32 has previously been shown to cause Bardet-Biedl syndrome \cite{chiang06}, a disease affecting multiple organs including the retina. 

The original data consists of 31,099 probe sets. For our analysis, we included only the 10,000 probe sets with the largest variances in expression (on the log scale). This resulted in $n = 120$ and $p = 10,000$. We then fit the model \eqref{linearregression} with an SSL penalty.  We compared the SSL approach to LASSO, SCAD, and MCP. 

To assess predictive accuracy, we randomly split the data set into 90 training observations and 30 test observations. We then fit the models on the training set and used the estimated  $\widehat{\boldsymbol{\beta}}_{\textrm{train}}$ to compute the mean squared prediction error (MSPE) on the left-out test set,
\begin{equation*} 
	\textrm{MSPE} = \frac{1}{30} \sum_{i=1}^{30} (y_{i, \textrm{test}} - \boldsymbol{x}_{i, \textrm{test}}^T \widehat{\boldsymbol{\beta}}_{\textrm{train}})^2,
\end{equation*}
where $(\boldsymbol{x}_{i, \textrm{test}}, y_{i, \textrm{test}}), i = 1, \ldots, 30,$ are the observations in the test set. We repeated this process 100 times and took the average MSPE. 

Table \ref{Table:2} shows the results for our analysis, as well as the number of selected probe sets when we fit the different models to the complete data set. SSL had the lowest out-of-sample MSPE, indicating the highest predictive power. MCP selected the most parsimonious model, with only nine probe sets out of the 10,000 selected. However, MCP also had a much higher MSPE than the other methods. On the other hand, SSL selected 28 probe sets (compared to 32 and 44 for LASSO and SCAD respectively) and still achieved the lowest MSPE. Our analysis illustrates that on this particular data set, SSL achieved both the best predictive performance and parsimony.

\begin{table}[t!]  
	\centering
	\begin{tabular}{l c c}
		\hline
		& MSPE & Number of selected probe sets \\ 
		\hline
		SSL & \textbf{0.011} & 28 \\
		LASSO & 0.012 & 32 \\
		SCAD & 0.015 & 44 \\ 
		MCP & 3.699 &  9 \\
		\hline
	\end{tabular}
	\caption{Average MSPE and the number of selected probe sets for the Bardet-Beidl Syndrome data analysis.}  
	\label{Table:2}
\end{table}

Of the 28 probe sets selected by SSL as being significantly associated with TRIM32, 14 of them had identifiable gene symbols. These genes were SCGB1A1, CELF1, ASXL3, FGFR2, MOBP, TGM7, SLC39A6, DDX58, TFF2, CLOCK, DUS4L, HTR5B, BIK, and SLC16A6.  In particular, according to \texttt{https://www.genecards.org} \footnote{Accessed from \texttt{https://www.genecards.org/cgi-bin/carddisp.pl?gene=TRIM32} on October 11, 2020.}, SCGB1A1 is known to be an interacting protein for the TRIM32 gene. The other associations that we found may be useful for researchers in studying the genetic factors contributing to Bardet-Biedl syndrome.

\section{Methodological extensions} \label{MethodologicalExtensions}

While we have focused on the normal linear regression model \eqref{linearregression} in Sections \ref{FrequentistVsBayes}-\ref{Illustration}, the spike-and-slab LASSO methodology has now been adopted in a variety of other statistical applications. In this section, we survey some of the extensions of the SSL to models beyond the normal linear regression framework.

\textbf{Generalized linear models (GLMs).} GLMs allow for a flexible generalization of the normal linear regression model \eqref{linearregression} which can accommodate categorical and count data, in addition to continuous variables. Letting $\boldsymbol{x}_i \in \mathbb{R}^{p}$ denote a vector of covariates for the $i$th observation, GLMs assume that the mean of the response variable is related to the linear predictor via a link function,
\begin{equation} \label{linkfunction}
	E(y_i \C \boldsymbol{x}_i) = h^{-1} ( \boldsymbol{x}_i^T \boldsymbol{\beta} ),
\end{equation}
and that the data distribution is expressed as
\begin{equation} \label{GLMlikelihood}
	p( \boldsymbol{y} \C \boldsymbol{X}, \boldsymbol{\beta}, \varphi ) = \prod_{i=1}^{n} p(y_i \C \boldsymbol{x}_i, \boldsymbol{\beta}, \varphi),
\end{equation}
where $\varphi$ is a dispersion parameter and the distribution $p(y_i \C \boldsymbol{x}_i, \boldsymbol{\beta}, \varphi )$ can take various forms, including normal, binomial and Poisson distributions. Obviously, the normal linear model \eqref{linearregression} is a special case of \eqref{linkfunction} with the identity link function $h(u)=u$.

\cite{TangShenZhangYi2017GLM} extended the SSL \eqref{SpikeAndSlabLASSO} to GLMs, including binary regression and Poisson regression, by placing the SSL prior \eqref{SSLprior} on the coefficients vector $\boldsymbol{\beta} \in \mathbb{R}^{p}$ in \eqref{linkfunction} and developing an EM algorithm to perform MAP estimation for $\boldsymbol{\beta}$. For inference with grouped variables in GLMs, \cite{TangShenLiZhangWenQianZhuangShiYi2018} further employed group-specific sparsity parameters $\theta_g$ for each group of variables, instead of a single $\theta$, as in \eqref{SpikeAndSlabLASSO}. 

\textbf{Survival analysis.} The SSL has also proven to be useful for predicting censored survival outcomes and detecting and estimating the effects of relevant covariates. Cox proportional hazards models are the most widely used method for studying the relationship between a censored survival response and an explanatory variable $\boldsymbol{x}_i \in \mathbb{R}^{p}$ \cite{KleinMoeschberger2003}. This model assumes that the hazard function of survival time $t$ takes the form,
\begin{equation} \label{CoxRegression}
	h(t \C \boldsymbol{x}_i ) = h_0(t) \exp(\boldsymbol{x}_i^T \boldsymbol{\beta}).
\end{equation}
\cite{TangShenZhangYi2017Cox} introduced the spike-and-slab LASSO Cox model which endows the coefficients $\boldsymbol{\beta}$ in \eqref{CoxRegression} with the SSL prior \eqref{SSLprior}. They developed an EM coordinate ascent algorithm to fit SSL Cox models. \cite{TangLeiZhangYiGuoChenShenYi2019} further introduced the GssLASSO Cox model which incorporates grouping information by endowing each group of coefficients with a group-specific sparsity parameter $\theta_g$ instead of the single $\theta$ of \eqref{SSLprior}.

\textbf{Sparse factor analysis and biclustering.} Factor models aim to explain the dependence structure among high-dimensional observations through a sparse decomposition of a $p \times p$ covariance matrix $\boldsymbol{\Omega}$ as $\boldsymbol{B} \boldsymbol{B}^T + \boldsymbol{\Sigma}$ where $\boldsymbol{B}$ is a $p \times K$ factor loadings matrix with $K \ll p$ and $\boldsymbol{\Sigma} = \textrm{diag}(\sigma_1^2, \ldots, \sigma_p^2)$. A generic latent factor model is
\begin{equation} \label{factormodel}
	\boldsymbol{y}_i = \boldsymbol{B} \boldsymbol{\eta}_{i} + \boldsymbol{\varepsilon}_i, \hspace{.5cm} \boldsymbol{\varepsilon}_i \sim \mathcal{N}_p (\boldsymbol{0}, \boldsymbol{\Sigma}),
\end{equation}
where $\boldsymbol{y}_i$ is a $p$-dimensional continuous response and $\boldsymbol{\eta}_i \sim \mathcal{N}_K (\boldsymbol{0}, \boldsymbol{I}_K)$ are unobserved latent factors. Many existing factor analysis approaches entail prespecification of the unknown factor cardinality $K$ and \textit{post hoc} rotations of the original solution to sparsity.  For the factor model \eqref{factormodel}, \cite{RockovaGeorge2016} endowed the entries of the loading matrix $\boldsymbol{B}$ with independent SSL priors,
\begin{align*}
	p(\beta_{jk} \C \gamma_{jk}, \lambda_{0k}, \lambda_1) = (1-\gamma_{jk}) \psi(\beta_{jk} \C \lambda_{0k} ) + \gamma_{jk}\psi(\beta_{jk}\C \lambda_{1}).
\end{align*}
However, instead of endowing each of the indicators $\gamma_{jk}$ with the usual beta-Bernoulli prior as in \eqref{SSLprior}, \cite{RockovaGeorge2016} endowed these with the Indian buffet process (IBP) prior \cite{GriffithsGhahramani2011}, which avoids the need to prespecify $K$. Further, \cite{RockovaGeorge2016} developed a parameter-expanded EM (PXL-EM) algorithm which employs \textit{intermediate} orthogonal rotations rather than post hoc rotations. In addition to obtaining a sparse solution, the PXL-EM algorithm also converges much faster than the vanilla EM algorithm and offers robustness against poor initializations.

For the problem of biclustering, i.e. identifying clusters using only subsets of their associated features, \cite{MoranRockovaGeorge2019} utilized the factor model \eqref{factormodel} in which \textit{both} the factors $\boldsymbol{\eta} = [ \boldsymbol{\eta}_1^T, \ldots, \boldsymbol{\eta}_n^T ] \in \mathbb{R}^{n \times K}$ and the loadings are sparse. To achieve a doubly sparse representation, \cite{MoranRockovaGeorge2019} placed an SSL prior coupled with an IBP prior on the factors and an SSL prior coupled  with a beta-Bernoulli prior on the loadings. An EM algorithm with a variational step was developed to implement spike-and-slab LASSO biclustering.

\textbf{Graphical models.} Suppose we are given data $\boldsymbol{Y} = (\boldsymbol{y}_1, \ldots, \boldsymbol{y}_n)^T$, where the $\boldsymbol{y}_i$'s are assumed to be iid $p$-variate random vectors distributed as $\mathcal{N}_p (\boldsymbol{0}, \boldsymbol{\Omega}^{-1})$ and $p>n$. In this setting, off-diagonal zero entries $\omega_{ij}$ encode conditional independence between variables $i$ and $j$.	To obtain a sparse estimate of $\boldsymbol{\Omega} = ( \omega_{i,j} )_{i,j}$, \cite{GanNarisettyLiang2018} introduced the following prior on $\boldsymbol{\Omega}$:
\begin{align} \label{graphicalmodelprior}
	p(\boldsymbol{\Omega}) = \prod_{i < j} \left[ (1-\theta) \psi(\omega_{ij}\C \lambda_0 ) + \theta \psi(\omega_{ij} \C \lambda_1 ) \right] \prod_{i=1}^{p} [ \tau e^{-\tau \omega_{ii}} ] \mathbb{I}( \Omega \succ 0 ) \mathbb{I}( \lVert \boldsymbol{\Omega} \rVert_2 \leq B ),
\end{align}
for some $\tau > 0, B > 0$. Here, $\boldsymbol{\Omega} \succ 0$ denotes that $\boldsymbol{\Omega}$ is positive-definite and $\lVert \boldsymbol{\Omega} \rVert_2$ denotes the spectral norm of $\boldsymbol{\Omega}$. The prior on $\boldsymbol{\Omega}$ \eqref{graphicalmodelprior} entails independent exponential priors on the diagonal entries and SSL priors on the off-diagonal entries. A similar prior formulation was considered in \cite{DeshpandeRockovaGeorge2019}, except \cite{DeshpandeRockovaGeorge2019} did not constrain $\boldsymbol{\Omega}$ to lie in the space of $p \times p$ matrices with uniformly bounded spectral norm.  \cite{GanNarisettyLiang2018} showed that constraining the parameter space for $\boldsymbol{\Omega}$ in such a way ensures that a) the corresponding optimization problem for the posterior mode is \textit{strictly} convex, and b) the posterior mode is a symmetric positive definite matrix. \cite{GanNarisettyLiang2018} developed an EM algorithm to estimate the posterior mode of $p(\boldsymbol{\Omega} \C \boldsymbol{Y})$.

\hspace{.2cm} The SSL prior was also extended to perform \textit{joint} estimation of multiple related Gaussian graphical models by \cite{LiMcCormickClark2019}. \cite{LiMcCormickClark2019} 
leveraged similarities in the underlying sparse precision matrices and developed an EM algorithm to perform this joint estimation.

\textbf{Seemingly unrelated regression models.} In seemingly unrelated regression models, multiple correlated responses are regressed on multiple predictors. The multivariate linear regression is an important case. Letting $\boldsymbol{y}_i \in \mathbb{R}^{q}$ be the vector of $q$ responses and $\boldsymbol{x}_i \in \mathbb{R}^{p}$ be the vector of $p$ covariates, this model is
\begin{equation} \label{multivariateregression}
	\boldsymbol{y}_i = \boldsymbol{x}_i^T \boldsymbol{B} + \boldsymbol{\varepsilon}_i, \hspace{.5cm} \boldsymbol{\varepsilon}_i \sim \mathcal{N}_q ( \boldsymbol{0}, \boldsymbol{\Omega}^{-1} ),
\end{equation}
\cite{DeshpandeRockovaGeorge2019} introduced the \textit{multivariate spike-and-slab LASSO} (mSSL) to perform joint selection and estimation from the $p \times q$ matrix of regressors $\boldsymbol{B}$ \textit{and} the precision matrix $\boldsymbol{\Omega}$. To obtain a sparse estimate of $(\boldsymbol{B}, \boldsymbol{\Omega})$, \cite{DeshpandeRockovaGeorge2019} placed the SSL prior \eqref{SSLprior} on the individual entries $\beta_{jk}, 1 \leq j \leq p, 1 \leq k \leq q$ in $\boldsymbol{B}$ and a product prior similar to \eqref{graphicalmodelprior} on $\boldsymbol{\Omega}$ (except \cite{DeshpandeRockovaGeorge2019} did not constrain $\boldsymbol{\Omega}$ to have bounded spectral norm). An expectation/conditional maximization (ECM) algorithm was developed to perform this joint estimation.

\textbf{Causal inference.} In observational studies, we are often interested in estimating the causal effect of a treatment $T$ on an outcome $y$, which requires proper adjustment of a set of potential confounders $\boldsymbol{x} \in \mathbb{R}^{p}$. When $p > n$, direct control for all potential confounders is infeasible and standard methods such as propensity scoring \cite{RosenbaumRubin1983} often fail. In this case, it is crucial to impose a low-dimensional structure on the confounder space. Given data $(y_i, T_i, \boldsymbol{x}_i), i=1, \ldots, n,$ where $T_i$ is the treatment effect,  \cite{AntonelliParmigianiDominici2019} estimated the (homogeneous) average treatment effect (ATE) $\Delta(t_1, t_2 ) = E(Y(t_1)-Y(t_2))$ by utilizing the model,
\begin{equation} \label{homogeneoustreatmentefectmodel}
	y_i \C T_i, \boldsymbol{x}_i, \beta_0, \beta_t, \boldsymbol{\beta}, \sigma^2 \sim \mathcal{N}(0, \beta_0 + \beta_t T_i+ \boldsymbol{x}_i ^T \boldsymbol{\beta}, \sigma^2 ).
\end{equation} 
Under \eqref{homogeneoustreatmentefectmodel}, the ATE is straightforwardly estimated as $\Delta (t_1, t_2) = (t_1 - t_2) \widehat{\beta}_t$. \cite{AntonelliParmigianiDominici2019} endowed the coefficients of the confounders $\boldsymbol{\beta}$ with the SSL prior \eqref{SSLprior}. In addition, \cite{AntonelliParmigianiDominici2019} also weighted the sparsity parameter $\theta$ in \eqref{SSLprior} by raising $\theta$ to a power $w_j, j=1, \ldots, p$, for each covariate, in order to better prioritize variables belonging to the slab (i.e. $\gamma_j = 1$) if they are also associated with the treatment. \cite{AntonelliParmigianiDominici2019} further extended the model \eqref{homogeneoustreatmentefectmodel} to the more general case of heterogeneous treatment effects.

\textbf{Regression with grouped variables.} Group structure arises in many statistical applications. For example, in genomics, genes within the same pathway may form a group and act in tandem to regulate a biological system. For regression with grouped variables, we can model the response $\boldsymbol{y}$ as
\begin{equation} \label{groupmodel}
	\boldsymbol{y} = \displaystyle \sum_{g=1}^{G} \boldsymbol{X}_g \boldsymbol{\beta}_g + \boldsymbol{\varepsilon}, \hspace{.5cm} \boldsymbol{\varepsilon} \sim \mathcal{N}_n ( \boldsymbol{0}, \sigma^2 \boldsymbol{I}_n),
\end{equation}
where $\boldsymbol{\beta}_g \in \mathbb{R}^{m_g}$ is a coefficients \textit{vector} of length $m_g$, and $\boldsymbol{X}_g$ is an $n \times m_g$ covariate matrix corresponding to group $g = 1, \ldots G$. Under model \eqref{groupmodel}, it is often of practical interest to select non-negligible groups and estimate their effects. To this end, \cite{BaiMoranAntonelliChenBoland2019} introduced the \textit{spike-and-slab group lasso} (SSGL). To regularize groups of coefficients, the SSGL replaces the univariate Laplace densities in the univariate SSL \eqref{SSLprior} with \textit{multivariate} Laplace densities. The SSGL prior is
\begin{equation} \label{ssgroupLASSO}
	\begin{array}{rl}
		p ( \boldsymbol{\beta} \C \theta ) = & \displaystyle \prod_{g=1}^{G} \left[ (1- \theta) \boldsymbol{\Psi} ( \boldsymbol{\beta}_g \C \lambda_0 ) + \theta \boldsymbol{\Psi} ( \boldsymbol{\beta}_g \C \lambda_1 ) \right],  \\
		\theta \sim & \mathcal{B}eta(a,b),
	\end{array}
\end{equation}
where $\boldsymbol{\Psi} (\boldsymbol{\beta}_g \C \lambda ) \propto \lambda^{m_g} e^{- \lambda \lVert \boldsymbol{\beta}_g \rVert_2 }$ and $\lambda_0 \gg \lambda_1$. The SSGL \eqref{ssgroupLASSO} is a two-group refinement of an $\ell_2$ penalty on groups of coefficients. Accordingly, the posterior mode under the SSGL thresholds entire groups of coefficients to zero, while simultaneously estimating the effects of nonzero groups and circumventing the estimation bias of the original group lasso \cite{YuanLin2006}. \cite{BaiMoranAntonelliChenBoland2019} developed an efficient blockwise-coordinate ascent algorithm to implement the SSGL model.

\textbf{Nonparameteric additive regression.} The advent of the SSGL prior \eqref{ssgroupLASSO} paved the way for the spike-and-slab lasso methodology to be extended to nonparametric problems. \cite{BaiMoranAntonelliChenBoland2019} introduced the \textit{nonparametric spike-and-slab lasso} (NPSSL) for sparse generalized additive models (GAMs). Under this model, the response surface is decomposed into the sum of univariate functions,
\begin{align} \label{GAM}
	y_i = \sum_{j=1}^{p} f_j(x_{ij}) + \varepsilon_i, \hspace{.5cm} \varepsilon_i \overset{iid}{\sim} \mathcal{N}(0, \sigma^2).
\end{align}
In \cite{BaiMoranAntonelliChenBoland2019}, each of the $f_j$'s is approximated using a basis expansion, or a linear combination of basis functions $\mathcal{B}_j = \{ g_{j1}, \ldots, g_{jd} \}$, i.e.
\begin{equation} \label{basisexpansion}
	f_j(x_{ij}) \approx \sum_{k=1}^{d} g_{jk} ( x_{ij}) \beta_{jk}.
\end{equation}
Under sparsity, most of the $f_j$'s in \eqref{GAM} are assumed to be $f_j = 0$. This is equivalent to assuming that most of the weight vectors $\boldsymbol{\beta}_j = (\beta_{j1}, \ldots, \beta_{jd})^T$, $j=1, \ldots, p$, in \eqref{basisexpansion} are equal to $\boldsymbol{0}_d$. The NPSSL is implemented by endowing the basis coefficients $\boldsymbol{\beta} = (\boldsymbol{\beta}_1^T, \ldots, \boldsymbol{\beta}_p^T)^T$ with the SSGL prior \eqref{ssgroupLASSO} to simultaneously select and estimate nonzero functionals. In addition, \cite{BaiMoranAntonelliChenBoland2019} also extended the NPSSL to identify and estimate the effects of nonlinear interaction terms $f_{rs} (X_{ir}, X_{is}), r \neq s$.

\textbf{Functional regression.} The spike-and-slab lasso methodology has also been extended to functional regression, where the response $y(t)$ is a function that \textit{varies} over some continuum $T$ (often time) A very popular model in this framework is the nonparametric varying coefficient model,
\begin{equation} \label{vcmodel}
	y_i(t) = \sum_{k=1}^{p} x_{ik}(t) \beta_k(t) + \varepsilon_i(t), \hspace{.5cm} t \in T,
\end{equation}
where $y_i(t)$ and $x_{ik}(t)$ are time-varying responses and covariates respectively and $\varepsilon_i(t)$ is a zero-mean stochastic process which captures the within-subject temporal correlations for the $i$th subject. Under \eqref{vcmodel}, the $\beta_k(t)$'s are smooth functions of time (possibly $\beta_k(t) = 0$ for all $t \in T$), and our primary interest is in estimation and variable selection from the $\beta_k(t)$'s, $k=1, \ldots, p$.

\hspace{.3cm} \cite{BaiBolandChen2019} introduced the nonparametric varying coefficient spike-and-slab lasso (NVC-SSL) to simultaneously select and estimate the smooth functions $\beta_k(t), k=1, \ldots, p$. Similarly as with the NPSSL, these functions are approximated using basis expansions of smoothing splines, and the basis coefficients are endowed with the SSGL prior \eqref{ssgroupLASSO}. Unlike GAMs, however, the NVC-SSL model does \textit{not} assume homoscedastic, independent error terms. Instead, the NVC-SSL model accounts for within-subject temporal correlations in its estimation procedure.  

\textbf{False discovery rate control with missing data}. Sorted L-One Penalized Estimator (SLOPE) is an elaboration of the LASSO tailored to false discovery control by assigning more penalty to the larger coefficients \cite{BodganVanDenBergSabattiSuCandes2015}.  SSL, on the other hand, penalizes large coefficients less and its false discovery rate is ultimately determined by a combination of the  prior inclusion weight $\theta$ and penalties $\lambda_1$ and $\lambda_0$. \cite{JiangBogdanJosseMiasojedowRockovaTBG2019} propose a hybrid procedure called adaptive Bayesian SLOPE, which effectively combines the SLOPE method (sorted $l_1$ regularization) together with the SSL method in the context of variable selection with missing covariate values. 
As with SSL, the coefficients are regarded as arising from a hierarchical model consisting of two groups: (1) the spike for the inactive and (2) the slab for the active. However, instead of assigning spike priors for each covariate, they propose a joint ``SLOPE'' spike prior which takes into account ordering of coefficient magnitudes in order to control for false discoveries.

\section{Theoretical properties} \label{SSLTheory}

In addition to its computational tractability and its excellent finite-sample performance, the spike-and-slab LASSO has also been shown to provide strong theoretical guarantees.  Although this paper focuses mainly on methodology, we briefly outline a few of the major theoretical developments for spike-and-slab LASSO methods.

A common theme in Bayesian asymptotic theory is the study of the learning rate of posterior point estimates (such as the mean, median or mode) and/or of the full posterior. Working under the frequentist assumption of a ``true'' underlying model, the aim under the former is to study the \textit{estimation} rate of point estimators under a given risk function, such as expected squared error loss. From a fully Bayes perspective, one may also be interested in the \textit{posterior contraction rate}, or the speed at which the \textit{entire} posterior contracts around the truth. In both cases, {  the frequentist minimax estimation rate is a useful benchmark, since the posterior cannot contract faster than this rate} \cite{GhosalGhoshVanDerVaart2000}.

In a variety of contexts, including the Gaussian sequence model, sparse linear regression, and graphical models, the SSL global posterior mode has been shown to achieve the {  minimax estimation rate} \cite{GanNarisettyLiang2018,Rockova2018,RockovaGeorge2016Abel,RockovaGeorge2018}. From a fully Bayesian perspective, the \textit{entire} posterior under SSL or SSL-type priors has \textit{also} been shown to achieve (near) optimal posterior contraction rates in the contexts of the Gaussian sequence model, linear regression, regression with grouped variables, nonparametric additive regression, and functional regression \cite{BaiBolandChen2019, BaiMoranAntonelliChenBoland2019, NieRockova2020, Rockova2018, RockovaGeorge2016Abel,RockovaGeorge2018}. It is not necessarily the case that the posterior mode and the full posterior contract at the same rate \cite{CastilloMismer2018,CastilloSchmidtHieberVanDerVaart2015}. These theoretical results thus show that the SSL is optimal from \textit{both} penalized likelihood \textit{and} fully Bayesian perspectives.

\section{Discussion} \label{Discussion}

In this paper, we have reviewed the spike-and-slab LASSO \eqref{SSLprior}. The SSL forms a continuum between the {  penalized likelihood LASSO and the Bayesian point-mass spike-and-slab frameworks}, borrowing strength from both constructs while addressing limitations of each. First, the SSL employs a \textit{non}-separable penalty that self-adapts to ensemble information about sparsity and that performs selective shrinkage. Second, the SSL is amenable to fast maximum \textit{a posteriori} finding algorithms which can be implemented in a highly efficient, scalable manner. Third, the posterior mode under the SSL prior can automatically perform both variable selection and estimation. Finally, uncertainty quantification for the SSL can be attained by either debiasing the posterior modal estimate or by utilizing efficient approaches to posterior sampling. Beyond linear regression, the spike-and-slab LASSO methodology is broadly applicable to a wide number of statistical problems, including generalized linear models, factor analysis, graphical models, and nonparametric regression. 

{ 
	\section*{Acknowledgments}
	This work was supported by funding from the University of South Carolina College of Arts \& Sciences, the James S. Kemper Foundation Faculty Research Fund at the University of Chicago Booth School of Business, and NSF Grants DMS-1916245, DMS-1944740.
}

\bibliographystyle{apa}
\bibliography{SSLReviewReferences}

\begin{thebibliography}{}

\bibitem[\protect\astroncite{Antonelli
  et~al.}{2019}]{AntonelliParmigianiDominici2019}
Antonelli, J., Parmigiani, G., and Dominici, F. (2019).
\newblock High-dimensional confounding adjustment using continuous spike and
  slab priors.
\newblock {\em Bayesian Analysis}, 14(3):805--828.

\bibitem[\protect\astroncite{Armagan et~al.}{2013}]{ArmaganDunsonLee2013}
Armagan, A., Dunson, D.~B., and Lee, J. (2013).
\newblock Generalized double pareto shrinkage.
\newblock {\em Statist. Sinica}, 23(1):119--143.

\bibitem[\protect\astroncite{Bai et~al.}{2020a}]{BaiBolandChen2019}
Bai, R., Boland, M.~R., and Chen, Y. (2020a).
\newblock Fast algorithms and theory for high-dimensional {B}ayesian varying
  coefficient models.
\newblock {\em arXiv pre-print arXiv: 1907.06477}.

\bibitem[\protect\astroncite{Bai and Ghosh}{2021}]{BaiGhosh2021}
Bai, R. and Ghosh, M. (2021).
\newblock On the beta prime prior for scale parameters in high-dimensional
  bayesian regression models.
\newblock {\em Statistica Sinica (to appear)}.

\bibitem[\protect\astroncite{Bai
  et~al.}{2020b}]{BaiMoranAntonelliChenBoland2019}
Bai, R., Moran, G.~E., Antonelli, J.~L., Chen, Y., and Boland, M.~R. (2020b).
\newblock Spike-and-slab group lassos for grouped regression and sparse
  generalized additive models.
\newblock {\em Journal of the American Statistical Association (to appear)}.

\bibitem[\protect\astroncite{Barbieri and Berger}{2004}]{BarbieriBerger2004}
Barbieri, M.~M. and Berger, J.~O. (2004).
\newblock Optimal predictive model selection.
\newblock {\em The Annals of Statistics}, 32(3):870--897.

\bibitem[\protect\astroncite{Belloni
  et~al.}{2011}]{BelloniChernozhukovWang2011}
Belloni, A., Chernozhukov, V., and Wang, L. (2011).
\newblock Square-root lasso: pivotal recovery of sparse signals via conic
  programming.
\newblock {\em Biometrika}, 98(4):791--806.

\bibitem[\protect\astroncite{Bhadra
  et~al.}{2019}]{BhadraDattaPolsonWillard2019}
Bhadra, A., Datta, J., Polson, N.~G., and Willard, B. (2019).
\newblock Lasso meets horseshoe: A survey.
\newblock {\em Statist. Science}, 34(3):405--427.

\bibitem[\protect\astroncite{Bhattacharya
  et~al.}{2016}]{BhattacharyaChakrabortyMallick2016}
Bhattacharya, A., Chakraborty, A., and Mallick, B.~K. (2016).
\newblock Fast sampling with {G}aussian scale mixture priors in
  high-dimensional regression.
\newblock {\em Biometrika}, 103(4):985--991.

\bibitem[\protect\astroncite{Bhattacharya
  et~al.}{2015}]{BhattacharyaPatiPillaiDunson2015}
Bhattacharya, A., Pati, D., Pillai, N.~S., and Dunson, D.~B. (2015).
\newblock Dirichlet-laplace priors for optimal shrinkage.
\newblock {\em Journal of the American Statistical Association},
  110(512):1479--1490.
\newblock PMID: 27019543.

\bibitem[\protect\astroncite{Bogdan
  et~al.}{2015}]{BodganVanDenBergSabattiSuCandes2015}
Bogdan, M., van~den Berg, E., Sabatti, C., Su, W., and Cand\'{e}s, E.~J.
  (2015).
\newblock {SLOPE}-adaptive variable selection via convex optimization.
\newblock {\em The Annals of Applied Statistics}, 9(3):1103--1140.

\bibitem[\protect\astroncite{Bottolo and
  Richardson}{2010}]{BottoloRichardson2010}
Bottolo, L. and Richardson, S. (2010).
\newblock Evolutionary stochastic search for {B}ayesian model exploration.
\newblock {\em Bayesian Analysis}, 5(3):583--618.

\bibitem[\protect\astroncite{C.~M.~Carvalho}{2010}]{CarvalhoPolsonScott2010}
C.~M.~Carvalho, N. G.~Polson, J. G.~S. (2010).
\newblock The horseshoe estimator for sparse signals.
\newblock {\em Biometrika}, 97(2):465--480.

\bibitem[\protect\astroncite{Carvalho et~al.}{2009}]{CarvalhoPolsonScott2009}
Carvalho, C.~M., Polson, N.~G., and Scott, J.~G. (2009).
\newblock Handling sparsity via the horseshoe.
\newblock In van Dyk, D. and Welling, M., editors, {\em Proceedings of the
  Twelth International Conference on Artificial Intelligence and Statistics},
  volume~5 of {\em Proceedings of Machine Learning Research}, pages 73--80,
  Hilton Clearwater Beach Resort, Clearwater Beach, Florida USA. PMLR.

\bibitem[\protect\astroncite{Castillo and Mismer}{2018}]{CastilloMismer2018}
Castillo, I. and Mismer, R. (2018).
\newblock Empirical {B}ayes analysis of spike and slab posterior distributions.
\newblock {\em Electronic Journal of Statistics}, 12(2):3953--4001.

\bibitem[\protect\astroncite{Castillo
  et~al.}{2015}]{CastilloSchmidtHieberVanDerVaart2015}
Castillo, I., Schmidt-Hieber, J., and van~der Vaart, A. (2015).
\newblock Bayesian linear regression with sparse priors.
\newblock {\em The Annals of Statistics}, 43(5):1986--2018.

\bibitem[\protect\astroncite{Chiang et~al.}{2006}]{chiang06}
Chiang, A.~P., Beck, J.~S., Yen, H.-J., Tayeh, M.~K., Scheetz, T.~E.,
  Swiderski, R.~E., Nishimura, D.~Y., Braun, T.~A., Kim, K.-Y.~A., Huang, J.,
  et~al. (2006).
\newblock Homozygosity mapping with {SNP} arrays identifies {TRIM}32, an {E3}
  ubiquitin ligase, as a {B}ardet--{B}iedl syndrome gene ({BBS}11).
\newblock {\em Proceedings of the National Academy of Sciences},
  103(16):6287--6292.

\bibitem[\protect\astroncite{Deshpande
  et~al.}{2019}]{DeshpandeRockovaGeorge2019}
Deshpande, S.~K., Ro\v{c}kov\'{a}, V., and George, E.~I. (2019).
\newblock Simultaneous variable and covariance selection with the multivariate
  spike-and-slab lasso.
\newblock {\em Journal of Computational and Graphical Statistics},
  28(4):921--931.

\bibitem[\protect\astroncite{Fan and Li}{2001}]{FanLi2001}
Fan, J. and Li, R. (2001).
\newblock Variable selection via nonconcave penalized likelihood and its oracle
  properties.
\newblock {\em Journal of the American Statistical Association},
  96(456):1348--1360.

\bibitem[\protect\astroncite{Friedman
  et~al.}{2007}]{FriedmanHastieTibshirani2007}
Friedman, J., Hastie, T., and Tibshirani, R. (2007).
\newblock {Sparse inverse covariance estimation with the graphical lasso}.
\newblock {\em Biostatistics}, 9(3):432--441.

\bibitem[\protect\astroncite{Friedman
  et~al.}{2010}]{FriedmanHastieTibshirani2010}
Friedman, J., Hastie, T., and Tibshirani, R. (2010).
\newblock Regularization paths for generalized linear models via coordinate
  descent.
\newblock {\em Journal of Statistical Software}, 33(1):1--22.

\bibitem[\protect\astroncite{Gan et~al.}{2018}]{GanNarisettyLiang2018}
Gan, L., Narisetty, N.~N., and Liang, F. (2018).
\newblock Bayesian regularization for graphical models with unequal shrinkage.
\newblock {\em Journal of the American Statistical Association},
  114:1218--1231.

\bibitem[\protect\astroncite{George and McCulloch}{1993}]{GeorgeMcCulloch1993}
George, E.~I. and McCulloch, R.~E. (1993).
\newblock Variable selection via {G}ibbs sampling.
\newblock {\em Journal of the American Statistical Association},
  88(423):881--889.

\bibitem[\protect\astroncite{Ghosal et~al.}{2000}]{GhosalGhoshVanDerVaart2000}
Ghosal, S., Ghosh, J.~K., and van~der Vaart, A.~W. (2000).
\newblock Convergence rates of posterior distributions.
\newblock {\em The Annals of Statistics}, 28(2):500--531.

\bibitem[\protect\astroncite{Ghosh
  et~al.}{2016}]{GhoshTangGhoshChakrabarti2016}
Ghosh, P., Tang, X., Ghosh, M., and Chakrabarti, A. (2016).
\newblock Asymptotic properties of bayes risk of a general class of shrinkage
  priors in multiple hypothesis testing under sparsity.
\newblock {\em Bayesian Analysis}, 11(3):753--796.

\bibitem[\protect\astroncite{Griffin and Brown}{2010}]{GriffinBrown2010}
Griffin, J.~E. and Brown, P.~J. (2010).
\newblock Inference with normal-gamma prior distributions in regression
  problems.
\newblock {\em Bayesian Analysis}, 5(1):171--188.

\bibitem[\protect\astroncite{Griffiths and
  Ghahramani}{2011}]{GriffithsGhahramani2011}
Griffiths, T.~L. and Ghahramani, Z. (2011).
\newblock The {I}ndian buffet process: An introduction and review.
\newblock {\em Journal fo Machine Learning Research}, 12:1185--1224.

\bibitem[\protect\astroncite{Hans et~al.}{2007}]{HansDobraWest2007}
Hans, C., Dobra, A., and West, M. (2007).
\newblock Shotgun stochastic search for ``large p'' regression.
\newblock {\em Journal of the American Statistical Association},
  102(478):507--516.

\bibitem[\protect\astroncite{Ishwaran and Rao}{2005}]{IshwaranRao2005}
Ishwaran, H. and Rao, J.~S. (2005).
\newblock Spike and slab variable selection: Frequentist and bayesian
  strategies.
\newblock {\em The Annals of Statistics}, 33(2):730--773.

\bibitem[\protect\astroncite{Javanmard and
  Montanari}{2018}]{JavanmardMontanari2018}
Javanmard, A. and Montanari, A. (2018).
\newblock Debiasing the lasso: Optimal sample size for gaussian designs.
\newblock {\em The Annals of Statistics}, 46(6A):2593--2622.

\bibitem[\protect\astroncite{Jiang
  et~al.}{2019}]{JiangBogdanJosseMiasojedowRockovaTBG2019}
Jiang, W., Bogdan, M., Josse, J., Miasojedow, B., Rockova, V., and {TraumaBase
  Group} (2019).
\newblock Adaptive bayesian slope -- high-dimensional model selection with
  missing values.
\newblock {\em arXiv pre-print arXiv: 1907.06477}.

\bibitem[\protect\astroncite{Johndrow
  et~al.}{2020}]{JohndrowOrensteinBhattacharya2020}
Johndrow, J., Orenstein, P., and Bhattacharya, A. (2020).
\newblock Scalable approximate {MCMC} algorithms for the horseshoe prior.
\newblock {\em Journal of Machine Learning Research}, 21(73):1--61.

\bibitem[\protect\astroncite{Johnstone and
  Silverman}{2004}]{JohnstoneSilverman2004}
Johnstone, I.~M. and Silverman, B.~W. (2004).
\newblock Needles and straw in haystacks: Empirical {B}ayes estimates of
  possibly sparse sequences.
\newblock {\em The Annals of Statistics}, 32(4):1594--1649.

\bibitem[\protect\astroncite{Johnstone and
  Silverman}{2005}]{JohnstoneSilverman2005}
Johnstone, I.~M. and Silverman, B.~W. (2005).
\newblock Empirical {B}ayes selection of wavelet thresholds.
\newblock {\em The Annals of Statistics}, 33(4):1700--1752.

\bibitem[\protect\astroncite{Kim and Gao}{2019}]{KimGao2019}
Kim, Y. and Gao, C. (2019).
\newblock Bayesian model selection with graph structured sparsity.
\newblock {\em arXiv preprint arXiv:1902.03316}.

\bibitem[\protect\astroncite{Klein and
  Moeschberger}{2003}]{KleinMoeschberger2003}
Klein, J.~P. and Moeschberger, M.~L. (2003).
\newblock {\em Survival Analysis Techniques for Censored and Truncated Data}.
\newblock Second edition.

\bibitem[\protect\astroncite{Li et~al.}{2019}]{LiMcCormickClark2019}
Li, Z., Mccormick, T., and Clark, S. (2019).
\newblock Bayesian joint spike-and-slab graphical lasso.
\newblock In Chaudhuri, K. and Salakhutdinov, R., editors, {\em Proceedings of
  the 36th International Conference on Machine Learning}, volume~97 of {\em
  Proceedings of Machine Learning Research}, pages 3877--3885, Long Beach,
  California, USA. PMLR.

\bibitem[\protect\astroncite{Matthews}{1975}]{Matthews1975}
Matthews, B. (1975).
\newblock Comparison of the predicted and observed secondary structure of t4
  phage lysozyme.
\newblock {\em Biochimica et Biophysica Acta (BBA) - Protein Structure},
  405(2):442--451.

\bibitem[\protect\astroncite{Mazumder
  et~al.}{2011}]{MazumderFriedmanHastie2011}
Mazumder, R., Friedman, J.~H., and Hastie, T. (2011).
\newblock Sparsenet: Coordinate descent with nonconvex penalties.
\newblock {\em Journal of the American Statistical Association},
  106(495):1125--1138.
\newblock PMID: 25580042.

\bibitem[\protect\astroncite{McLachlan and
  Basford}{1988}]{McLachlanBasford1988}
McLachlan, G.~J. and Basford, K.~E. (1988).
\newblock {\em {M}ixture models: {I}nference and applications to clustering.}
\newblock Marcel Dekker, New York.

\bibitem[\protect\astroncite{Meinshausen and
  B{\"u}hlmann}{2006}]{MeinshausenBuhlmann2006}
Meinshausen, N. and B{\"u}hlmann, P. (2006).
\newblock High-dimensional graphs and variable selection with the lasso.
\newblock {\em The Annals of Statistics}, 34(3):1436--1462.

\bibitem[\protect\astroncite{Mitchell and
  Beauchamp}{1988}]{MitchellBeauchamp1988}
Mitchell, T. and Beauchamp, J. (1988).
\newblock Bayesian variable selection in linear regression.
\newblock {\em Journal of the American Statistical Association},
  83(404):1023--1032.

\bibitem[\protect\astroncite{Moran et~al.}{2020}]{MoranRockovaGeorge2019}
Moran, G.~E., Ro\u{c}kov\'{a}, V., and George, E.~I. (2020).
\newblock Spike-and-slab lasso biclustering.
\newblock {\em The Annals of Applied Statistics (to appear)}.

\bibitem[\protect\astroncite{Moran et~al.}{2019}]{MoranRockovaGeorge2018}
Moran, G.~E., Ro\v{c}kov\'{a}, V., and George, E.~I. (2019).
\newblock Variance prior forms for high-dimensional {B}ayesian variable
  selection.
\newblock {\em Bayesian Analysis}, 14(4):1091--1119.

\bibitem[\protect\astroncite{Narisetty and He}{2014}]{NarisettyHe2014}
Narisetty, N.~N. and He, X. (2014).
\newblock Bayesian variable selection with shrinking and diffusing priors.
\newblock {\em The Annals of Statistics}, 42(2):789--817.

\bibitem[\protect\astroncite{Narisetty et~al.}{2019}]{NarisettyShenHe2019}
Narisetty, N.~N., Shen, J., and He, X. (2019).
\newblock Skinny {G}ibbs: A consistent and scalable {G}ibbs sampler for model
  selection.
\newblock {\em Journal of the American Statistical Association},
  114(527):1205--1217.

\bibitem[\protect\astroncite{Nie and Ro\v{c}kov\'{a}}{2020}]{NieRockova2020}
Nie, L. and Ro\v{c}kov\'{a}, V. (2020).
\newblock Fast posterior sampling for the spike-and-slab {LASSO}.
\newblock {\em arXiv pre-print arXiv: 2011.14279}.

\bibitem[\protect\astroncite{Park and Casella}{2008}]{ParkCasella2008}
Park, T. and Casella, G. (2008).
\newblock The {B}ayesian lasso.
\newblock {\em Journal of the American Statistical Association},
  103(482):681--686.

\bibitem[\protect\astroncite{Polson and Sun}{2019}]{PolsonSun2019}
Polson, N.~G. and Sun, L. (2019).
\newblock Bayesian $\ell_0$-regularized least squares.
\newblock {\em Applied Stochastic Models in Business and Industry},
  35(3):717--731.

\bibitem[\protect\astroncite{Ray and Szab\'{o}}{2020}]{RaySzabo2020}
Ray, K. and Szab\'{o}, B. (2020).
\newblock Variational {B}ayes for high-dimensional linear regression with
  sparse priors.
\newblock {\em Journal of the American Statistical Association (to appear)}.

\bibitem[\protect\astroncite{Rosenbaum and Rubin}{1983}]{RosenbaumRubin1983}
Rosenbaum, P.~R. and Rubin, D.~B. (1983).
\newblock {The central role of the propensity score in observational studies
  for causal effects}.
\newblock {\em Biometrika}, 70(1):41--55.

\bibitem[\protect\astroncite{Ro\v{c}kov\'{a}}{2018}]{Rockova2018}
Ro\v{c}kov\'{a}, V. (2018).
\newblock Bayesian estimation of sparse signals with a continuous
  spike-and-slab prior.
\newblock {\em The Annals of Statistics}, 46(1):401--437.

\bibitem[\protect\astroncite{Ro\v{c}kov\'{a} and
  George}{2014}]{RockovaGeorge2014}
Ro\v{c}kov\'{a}, V. and George, E.~I. (2014).
\newblock \uppercase{EMVS}: The \uppercase{EM} approach to {B}ayesian variable
  selection.
\newblock {\em Journal of the American Statistical Association},
  109(506):828--846.

\bibitem[\protect\astroncite{Ro\v{c}kov\'{a} and
  George}{2016a}]{RockovaGeorge2016Abel}
Ro\v{c}kov\'{a}, V. and George, E.~I. (2016a).
\newblock Bayesian penalty mixing: The case of a non-separable penalty.
\newblock In {\em Statistical Analysis for High-Dimensional Data - The Abel
  Symposium 2014}, pages 233--254. Springer.

\bibitem[\protect\astroncite{Ro\v{c}kov\'{a} and
  George}{2016b}]{RockovaGeorge2016}
Ro\v{c}kov\'{a}, V. and George, E.~I. (2016b).
\newblock Fast {B}ayesian factor analysis via automatic rotations to sparsity.
\newblock {\em Journal of the American Statistical Association},
  111(516):1608--1622.

\bibitem[\protect\astroncite{Ro\v{c}kov\'{a} and
  George}{2018}]{RockovaGeorge2018}
Ro\v{c}kov\'{a}, V. and George, E.~I. (2018).
\newblock The spike-and-slab {LASSO}.
\newblock {\em Journal of the American Statistical Association},
  113(521):431--444.

\bibitem[\protect\astroncite{Ro\v{c}kov\'{a} and Moran}{2018}]{SSLASSOpackage}
Ro\v{c}kov\'{a}, V. and Moran, G.~E. (2018).
\newblock {\em \texttt{\uppercase{SSLASSO}}: The Spike-and-Slab
  \uppercase{LASSO}}.
\newblock R package version 1.2-1.

\bibitem[\protect\astroncite{Scheetz et~al.}{2006}]{Scheetz06}
Scheetz, T.~E., Kim, K.-Y.~A., Swiderski, R.~E., Philp, A.~R., Braun, T.~A.,
  Knudtson, K.~L., Dorrance, A.~M., DiBona, G.~F., Huang, J., Casavant, T.~L.,
  et~al. (2006).
\newblock Regulation of gene expression in the mammalian eye and its relevance
  to eye disease.
\newblock {\em Proceedings of the National Academy of Sciences},
  103(39):14429--14434.

\bibitem[\protect\astroncite{Scott and Berger}{2010}]{ScottBerger2010}
Scott, J.~G. and Berger, J.~O. (2010).
\newblock Bayes and empirical-{B}ayes multiplicity adjustment in the
  variable-selection problem.
\newblock {\em The Annals of Statistics}, 38(5):2587--2619.

\bibitem[\protect\astroncite{Sun and Zhang}{2012}]{SunZhang2012}
Sun, T. and Zhang, C.-H. (2012).
\newblock Scaled sparse linear regression.
\newblock {\em Biometrika}, 99(4):879--898.

\bibitem[\protect\astroncite{Tang
  et~al.}{2019}]{TangLeiZhangYiGuoChenShenYi2019}
Tang, Z., Lei, S., Zhang, X., Yi, Z., Guo, B., Chen, J.~Y., Shen, Y., and Yi,
  N. (2019).
\newblock Gsslasso {C}ox: a {B}ayesian hierarchical model for predicting
  survival and detecting associating genes by incorporating pathway
  information.
\newblock {\em BMC Bioinformatics}, 20(1).

\bibitem[\protect\astroncite{Tang
  et~al.}{2018}]{TangShenLiZhangWenQianZhuangShiYi2018}
Tang, Z., Shen, Y., Li, Y., Zhang, X., Wen, J., Qian, C., Zhuang, W., Shi, X.,
  and Yi, N. (2018).
\newblock Group spike-and-slab lasso generalized linear models for disease
  prediction and associated genes detection by incorporating pathway
  information.
\newblock {\em Bioinformatics}, 34(6):901--910.

\bibitem[\protect\astroncite{Tang et~al.}{2017a}]{TangShenZhangYi2017GLM}
Tang, Z., Shen, Y., Zhang, X., and Yi, N. (2017a).
\newblock The spike-and-slab lasso generalized linear models for prediction and
  associated genes detection.
\newblock {\em Genetics}, 205(1):77--88.

\bibitem[\protect\astroncite{Tang et~al.}{2017b}]{TangShenZhangYi2017Cox}
Tang, Z., Shen, Y., Zhang, X., and Yi, N. (2017b).
\newblock {The spike-and-slab lasso Cox model for survival prediction and
  associated genes detection}.
\newblock {\em Bioinformatics}, 33(18):2799--2807.

\bibitem[\protect\astroncite{Tibshirani}{1996}]{Tibshirani1996}
Tibshirani, R. (1996).
\newblock Regression shrinkage and selection via the lasso.
\newblock {\em Journal of the Royal Statistical Society: Series B (Statistical
  Methodology)}, 58:267--288.

\bibitem[\protect\astroncite{Ueda and Nakano}{1998}]{UedaNakano1998}
Ueda, N. and Nakano, R. (1998).
\newblock Deterministic annealing \uppercase{EM} algorithm.
\newblock {\em Neural Networks}, 11(2):271 -- 282.

\bibitem[\protect\astroncite{van~de Geer
  et~al.}{2014}]{VanDeGeerBuhlmannRitovDezeure2014}
van~de Geer, S., B{\"u}hlmann, P., Ritov, Y., and Dezeure, R. (2014).
\newblock On asymptotically optimal confidence regions and tests for
  high-dimensional models.
\newblock {\em The Annals of Statistics}, 42(3):1166--1202.

\bibitem[\protect\astroncite{Wellcome\:Trust}{2007}]{WellcomeTrust2007}
Wellcome\:Trust (2007).
\newblock Genome-wide association study of 14,000 cases of seven common
  diseases and 3000 shared controls.
\newblock {\em Nature}, 447:661--678.

\bibitem[\protect\astroncite{Yuan and Lin}{2006}]{YuanLin2006}
Yuan, M. and Lin, Y. (2006).
\newblock Model selection and estimation in regression with grouped variables.
\newblock {\em Journal of the Royal Statistical Society: Series B (Statistical
  Methodology)}, 68(1):49--67.

\bibitem[\protect\astroncite{Zhang}{2010}]{Zhang2010}
Zhang, C.-H. (2010).
\newblock Nearly unbiased variable selection under minimax concave penalty.
\newblock {\em The Annals of Statistics}, 38(2):894--942.

\bibitem[\protect\astroncite{Zhang and Zhang}{2014}]{ZhangZhang2014}
Zhang, C.-H. and Zhang, S.~S. (2014).
\newblock Confidence intervals for low dimensional parameters in high
  dimensional linear models.
\newblock {\em Journal of the Royal Statistical Society: Series B (Statistical
  Methodology)}, 76(1):217--242.

\bibitem[\protect\astroncite{Zhang and Zhang}{2012}]{ZhangZhang2012}
Zhang, C.-H. and Zhang, T. (2012).
\newblock A general theory of concave regularization for high-dimensional
  sparse estimation problems.
\newblock {\em Statistical Science}, 27(4):576--593.

\bibitem[\protect\astroncite{Zou}{2006}]{Zou2006}
Zou, H. (2006).
\newblock The adaptive lasso and its oracle properties.
\newblock {\em Journal of the American Statistical Association},
  101(476):1418--1429.

\bibitem[\protect\astroncite{Zou and Hastie}{2005}]{ZouHastie2005}
Zou, H. and Hastie, T. (2005).
\newblock Regularization and variable selection via the elastic net.
\newblock {\em Journal of the Royal Statistical Society: Series B (Statistical
  Methodology)}, 67(2):301--320.

\end{thebibliography}

\end{document}